\documentclass[journal,twocolumn]{IEEEtran}
\ifCLASSINFOpdf
\else
\fi
\usepackage{amsbsy,amscd,amsfonts,amsgen,amsmath,amsopn,amssymb,amstext,amsthm,amsxtra}
\usepackage{graphicx}\usepackage{subfigure}
\usepackage{verbatim}
\usepackage{citesort}
\usepackage{url}
\usepackage{color}
\usepackage{epstopdf}
\usepackage{array}
\usepackage{caption}
\usepackage{changes}
\usepackage{bbm}
\usepackage{algorithm}
\usepackage{algpseudocode}
\graphicspath{{./} {img/}}

\IEEEoverridecommandlockouts

\newcounter{cnt01}
\newcounter{cnt02}
\newcounter{cnt03}
\newcounter{cnt04}
\addtocounter{cnt01}{2}
\addtocounter{cnt02}{5}
\addtocounter{cnt03}{6}
\addtocounter{cnt04}{9}

\begin{document}
\title{Compressive Hyperspectral Imaging\\ via Approximate Message Passing}
\author{Jin~Tan,~\IEEEmembership{Student Member,~IEEE},
Yanting~Ma,~\IEEEmembership{Student Member,~IEEE},
Hoover Rueda,~\IEEEmembership{Student Member,~IEEE},
Dror~Baron,~\IEEEmembership{Senior Member,~IEEE},
and Gonzalo R. Arce,~\IEEEmembership{Fellow,~IEEE}
\thanks{Portions of the work will be presented at the IEEE Global Conf. Signal Inf. Process., Orlando, FL, December 2015.}
\thanks{The work of Jin Tan, Yanting Ma, and Dror Baron was supported in part by the National Science Foundation under the Grant CCF-1217749 and in part by the U.S. Army Research Office under the Contract W911NF-14-1-0314. The work of Hoover Rueda and Gonzalo R. Arce was supported by the U.S. Army Research Office under the Contract W911NF-12-1-0380.
}
\thanks{Jin Tan, Yanting Ma, and Dror Baron are with the Department of Electrical and Computer Engineering, NC State University, Raleigh, NC 27695. E-mail: \{jtan, yma7, barondror\}@ncsu.edu.
Hoover Rueda and Gonzalo R. Arce are with the Department of Electrical and Computer Engineering, University of Delaware, Newark, DE 19716. E-mail: \{rueda,arce\}@udel.edu.}
}

%
\maketitle
\thispagestyle{empty}
\newcommand{\xhat}{\widehat{\mathbf{x}}}
\newcommand{\xhati}{\widehat{x}_i}
\def\x{{\mathbf x}}
\def\L{{\cal L}}
\begin{abstract}
We consider a compressive hyperspectral imaging reconstruction problem, where three-dimensional spatio-spectral information about a scene is sensed by a coded aperture snapshot spectral imager (CASSI). The CASSI imaging process can be modeled as suppressing three-dimensional coded and shifted voxels and projecting these onto a two-dimensional plane, such that the number of acquired measurements is greatly reduced. On the other hand, because the measurements are highly compressive, the reconstruction process becomes challenging.
We previously proposed a compressive imaging reconstruction algorithm that is applied to two-dimensional images based on the approximate message passing (AMP) framework. AMP is an iterative algorithm that can be used in signal and image reconstruction by performing denoising at each iteration. We employed an adaptive Wiener filter as the image denoiser, and called our algorithm ``AMP-Wiener." 
In this paper, we extend AMP-Wiener to three-dimensional hyperspectral image reconstruction, and call it ``AMP-3D-Wiener."
Applying the AMP framework to the CASSI system is challenging, because the matrix that models the CASSI system is highly sparse, and such a matrix is not suitable to AMP and makes it difficult for AMP 
to converge. 
Therefore, we modify the adaptive Wiener filter and employ a technique called damping to solve for the divergence issue of AMP.
Our approach is applied in nature, and the numerical experiments
show that AMP-3D-Wiener 
outperforms existing widely-used algorithms such as gradient projection for sparse reconstruction (GPSR) and two-step iterative shrinkage/thresholding (TwIST) given a similar amount of runtime. Moreover, in contrast to GPSR and TwIST, AMP-3D-Wiener need not tune any parameters, which simplifies the reconstruction process. 
\end{abstract}
\IEEEpeerreviewmaketitle

\section{Introduction}
\label{sec:intro}
\subsection{Motivation}
\label{subsec:motivation}

A hyperspectral image is a three-dimensional (3D) image cube comprised of a collection of two-dimensional (2D)
images (slices), where
each 2D image is captured at a specific wavelength. 
Hyperspectral images allow us to analyze spectral information about each spatial point in a scene, and thus can help us identify different materials that appear 
in the scene~\cite{Heinz2001}. Therefore, hyperspectral imaging has applications 
to areas such as  medical imaging~\cite{Schultz2001,Panasyuk2007}, remote sensing~\cite{Schaepman2009}, geology~\cite{Kruse2003},
and astronomy~\cite{Hege2004}. 

Conventional spectral imagers include whisk broom scanners, push broom scanners~\cite{Brady2009,Eismann2012}, and spectrometers~\cite{Gat2000}. 
In whisk broom scanners, a mirror reflects light onto a single detector, so that one pixel of data is collected at a time;
in push broom scanners, an image cube is captured with one focal plane array (FPA) measurement per spatial line of the scene; and in spectrometers, a set of optical bandpass filters are tuned in steps in order to scan the scene.
The disadvantages of these techniques are that
({\em i}) data acquisition takes a long time, because they require scanning a number of zones linearly in proportion to the desired spatial and spectral resolution; and
({\em ii}) large amounts of data are acquired and must be stored and transmitted.
For example, for a megapixel camera ($10^6$ pixels) that captures a few hundred spectral bands ($>100$ spectral channels) at 8 or 16 bits per frame, conventional spectral imagers demand roughly 10 megabytes per raw spectral image, and thus require space on the order of gigabytes for transmission or storage, which exceeds existing streaming capabilities.

To address the limitations of conventional spectral imaging techniques, many spectral imager sampling schemes based on compressive sensing~\cite{DonohoCS,CandesRUP,BaraniukCS2007} have been proposed~\cite{Gehm2007,Yuan2015Side,August2013Hyper}. 
The coded aperture snapshot 
spectral imager (CASSI)~\cite{Gehm2007,Wagadarikar2008,Arguello2011,Wagadarikar2008single} is a popular compressive spectral imager and acquires
image data from different wavelengths simultaneously. 
In CASSI, the voxels of a scene are first coded by an aperture, then dispersed by a dispersive element, and finally detected by a 2D FPA. That is, a 3D image cube is suppressed and measured by a 2D array, and thus CASSI acquires far fewer measurements than those acquired by conventional spectral imagers, which significantly accelerates the imaging process. 
In particular, for a data cube with spatial resolution of $N\times M$ and $L$ spectral bands, conventional spectral imagers collect $MNL$ measurements. In contrast, CASSI collects measurements on the order of $M(N+L-1)$. Therefore, the acquisition time, storage space, and required bandwidth for transmission in CASSI are reduced.
On the other hand, because
the measurements from CASSI are highly compressive, reconstructing 3D image cubes from CASSI measurements becomes challenging.
Moreover, because of the massive size of 3D image data, it is desirable to develop fast reconstruction algorithms in order to realize real time acquisition and processing.

Fortunately, it is possible to reconstruct the 3D cube from the 2D measurements according to the theory of compressive sensing~\cite{DonohoCS,CandesRUP,BaraniukCS2007}, because the 2D images from different wavelengths are highly correlated, and the 3D image cube is sparse in an appropriate transform domain, meaning that only a small portion of the transform coefficients have large values.
Approximate message passing (AMP)~\cite{DMM2009} has recently become a popular algorithm that solves compressive sensing problems, owing to its promising performance and efficiency. Therefore, we are motivated to investigate how to apply AMP to the CASSI system.

\subsection{Related work}
\label{subsec:relatedWork}
Several algorithms have been
proposed to reconstruct image cubes from measurements acquired by CASSI. 
First, the reconstruction problem for the CASSI system can be solved by $\ell_1$-minimization. In Arguello and Arce~\cite{Arguello2014}, gradient projection for sparse reconstruction (GPSR)~\cite{GPSR2007} is utilized to solve for the $\ell_1$-minimization problem, where the sparsifying transform is the Kronecker product of a 2D wavelet transform and a 1D discrete cosine transform (DCT).
Besides using $\ell_1$-norm as the regularizer, total variation is a popular alternative;
Wagadarikar et al.~\cite{Wagadarikar2008} employed total variation~\cite{Chambolle2004,Chan2005} as the regularizer in the
two-step iterative shrinkage/thresholding (TwIST) framework~\cite{NewTWIST2007}, a modified and fast version of standard iterative shrinkage/thresholding. 
Apart from using the wavelet-DCT basis, 
one can 
sparsify image cubes by dictionary learning~\cite{Yuan2015Side}, or using Gaussian mixture models~\cite{Rajwade2013}.
An interesting idea to improve the reconstruction quality of the dictionary learning based approach is to use a standard image with red, green, and blue (RGB) components of the same scene as side information~\cite{Yuan2015Side}. That is, a coupled dictionary is learned from the joint datasets of the CASSI measurements and the corresponding RGB image.
We note in passing that using color sensitive RGB detectors directly as the FPA of CASSI is another way to improve the sensing of spectral images, because spatio-spectral coding can be attained in a single snapshot without requiring extra optical elements~\cite{Rueda2015}.

Despite the good results attained with the algorithms mentioned above, they all need manual tuning of some parameters, which may be time consuming. In GPSR and TwIST, 
the optimal regularization parameter could be different in reconstructing different image cubes.
In dictionary learning methods, although the parameters can be learned automatically by methods such as Markov Chain Monte Carlo, the  learning process is usually time consuming. Moreover,
the patch size and the number of dictionary atoms in dictionary learning methods must be chosen carefully.

\subsection{Contributions}
\label{sec:contrib}

In this paper, we develop a robust and fast reconstruction algorithm for the CASSI system using approximate message passing (AMP)~\cite{DMM2009}. AMP is an iterative algorithm that can apply image denoising at each iteration.
Previously, we proposed a 2D compressive imaging reconstruction algorithm, AMP-Wiener~\cite{Tan_CompressiveImage2014}, where an adaptive Wiener filter was applied as the image denoiser within AMP.
Our numerical results showed that AMP-Wiener outperformed the prior art
in terms of both reconstruction quality and runtime. 
The current paper extends AMP-Wiener to reconstruct 3D hyperspectral images from the CASSI system, and we call the new approach ``AMP-3D-Wiener."
Because the matrix that models the CASSI system is highly sparse, structured, and ill-conditioned,
applying AMP to the CASSI system becomes challenging. 
For example, ({\em i}) the noisy image cube that is obtained at each AMP iteration contains non-Gaussian noise; and ({\em ii}) AMP encounters divergence problems, i.e., the reconstruction error may increase with more iterations.
Although it is favorable to use a high-quality denoiser within AMP, so that the reconstruction error may decrease faster as the number of iteration increases, we have found that in such an ill-conditioned imaging system, applying aggressive denoisers within AMP causes divergence problems.
Therefore, besides using standard techniques such as damping ~\cite{Rangan2014ISIT,Vila2014} to encourage the convergence of AMP, we modify the adaptive Wiener filter and make it robust to the ill-conditioned system model. 
There are existing denoisers that may outperform the modified adaptive Wiener filter in a single step denoising problem. However, the modified adaptive Wiener filter fits into the AMP framework and allows AMP to improve over successive iterations.

Our approach is applied in nature, and the convergence of AMP-3D-Wiener is tested numerically.
We simulate AMP-3D-Wiener on several settings where complementary random coded apertures (see details in Section IV-A) are employed. The numerical results show that AMP-3D-Wiener reconstructs 3D image cubes with less runtime and higher quality than other compressive hyperspectral imaging reconstruction algorithms such as GPSR~\cite{GPSR2007} and TwIST~\cite{Wagadarikar2008,NewTWIST2007} (Figure~\ref{fig.iter_Psnr}), even when the regularization parameters in GPSR and TwIST have already been tuned. 
These favorable results provide AMP-3D-Wiener major advantages over GPSR and TwIST.
First,
when the bottleneck is the time required to run the reconstruction algorithm,
AMP-3D-Wiener can provide the same reconstruction quality in 100 seconds that the other algorithms provide in 450 seconds (Figure~\ref{fig.iter_Psnr}).
Second, when the bottleneck is the time required for signal acquisition by CASSI hardware, the improved reconstruction quality could allow to reduce the number of shots taken by CASSI by as much as a factor of $2$ (Figure~\ref{fig.shots_Psnr}).
Finally, the reconstructed image cube can be obtained by running AMP-3D-Wiener only once, because AMP-3D-Wiener does not need to tune any parameters. In contrast, the regularization parameters in GPSR and TwIST need to be tuned carefully, because the optimal values of these parameters may vary for different test image cubes. In order to tune the parameters for each test image cube, we run GPSR and TwIST many times with different parameter values, and then select the ones that provide the best results.

The remainder of the paper is arranged as follows. We review 
CASSI in Section~\ref{sec:CASSI}, and describe our AMP based compressive hyperspectral imaging reconstruction algorithm in Section \ref{sec:Algo}.
Numerical results are presented in Section \ref{sec:NumSim}, while Section~\ref{sec:disc} concludes.

\section{Coded Aperture Snapshot Spectral Imager (CASSI)}
\label{sec:CASSI}

\subsection{Mathematical representation of CASSI}
\label{subsec:MathCASSI}

The coded aperture snapshot spectral imager (CASSI)~\cite{Wagadarikar2008single} is a compressive spectral imaging system 
that collects far fewer measurements than traditional spectrometers. In CASSI, ({\em i}) the 2D spatial information of a scene is coded by an aperture, ({\em ii}) the coded spatial projections are spectrally shifted by a dispersive element, and ({\em iii}) the coded and shifted projections are detected by a 2D FPA. That is, in each coordinate of the FPA, the received projection is an integration of the coded and shifted voxels over all spectral bands at the same spatial coordinate. More specifically, let $f_0(x,y,\lambda)$ denote the voxel intensity of a scene at spatial coordinate $(x,y)$ and at wavelength $\lambda$, and let $T(x,y)$ denote the coded aperture. The coded density $T(x,y)f_0(x,y,\lambda)$ is then spectrally shifted by the dispersive element along one of the spatial dimensions. The energy received by the FPA at coordinate $(x,y)$ is therefore
\begin{equation}
g(x,y) = \int_\Lambda T(x,y-S(\lambda))f_0(x,y-S(\lambda),\lambda)d\lambda,
\label{eq:contInt}
\end{equation}
where $S(\lambda)$ is the dispersion function induced by the prism at wavelength $\lambda$. Suppose we take a scene of spatial dimension $M$ by $N$ and spectral dimension $L$, i.e., the dimension of the image cube is $M\times N\times L$, and the dispersion is along the second spatial dimension $y$, then the number of measurements captured by the FPA will be $M(N+L-1)$. If we approximate the integral in~\eqref{eq:contInt} by a discrete summation and vectorize the 3D image cube and the 2D measurements, then we obtain a matrix-vector form of~\eqref{eq:contInt},
\begin{equation}
{\bf g} = {\bf H}{\bf f_0}+{\bf z},
\label{eq:CASSI}
\end{equation}
where ${\bf f_0}$ is the vectorized 3D image cube of dimension $n=MNL$, vectors ${\bf g}$ and ${\bf z}$ are the measurements and the additive noise, respectively,
and the matrix ${\bf H}$ is an equivalent linear operator that models the integral in~\eqref{eq:contInt}.
In this paper, we assume that the additive noise ${\bf z}$ is independent and identically distributed (i.i.d.) Gaussian.
With a single shot of CASSI, the number of measurements is~$m=M (N+L-1)$, whereas $K$ shots will yield $m=K M(N+L-1)$ measurements.
The matrix H in~\eqref{eq:CASSI} accounts for the effects of the coded aperture and the dispersive element. A sketch of this matrix is depicted in Figure~\ref{fig.1a} when $K=2$ shots are used. It consists of a set of diagonal patterns that repeat in the horizontal direction, each time with a unit downward shift, as many times as the number of spectral bands. Each diagonal pattern is the coded aperture itself after being column-wise vectorized. Just below, the next set of diagonal patterns are determined by the coded aperture pattern used in the subsequent shot. The matrix H will thus have as many sets of diagonal patterns as FPA measurements.
Although ${\bf H}$ is sparse and highly structured, the restricted isometry property~\cite{Candes05a} still holds, as shown by Arguello and Arce~\cite{Arguello2012}.

\begin{figure}[t]
\vspace*{-5mm}
\subfigure[The matrix ${\bf H}$ for standard CASSI]{
	\includegraphics[width=80mm]{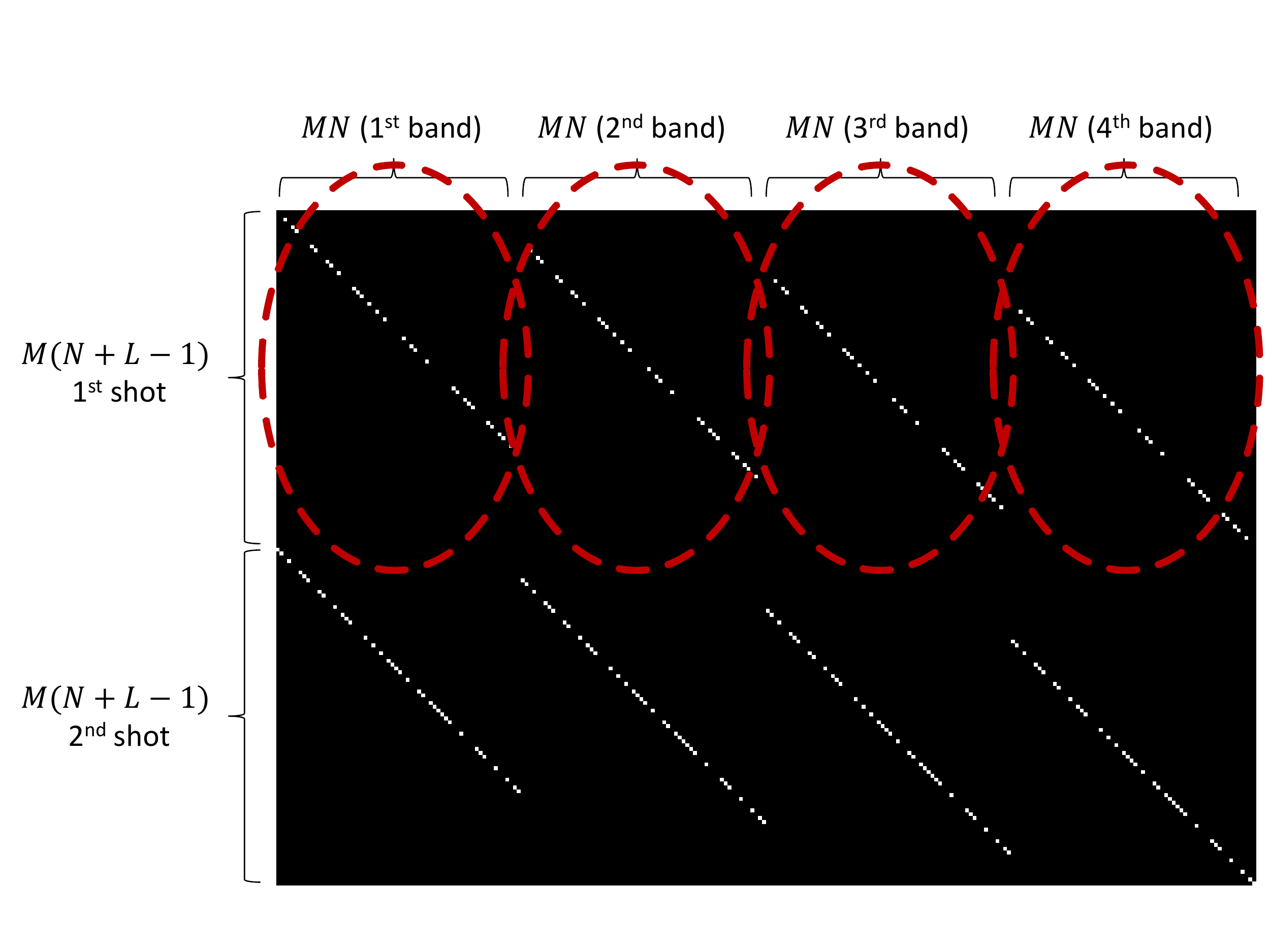}
	\label{fig.1a}
}
\subfigure[The matrix ${\bf H}$ for higher order CASSI]{
	\includegraphics[width=89mm]{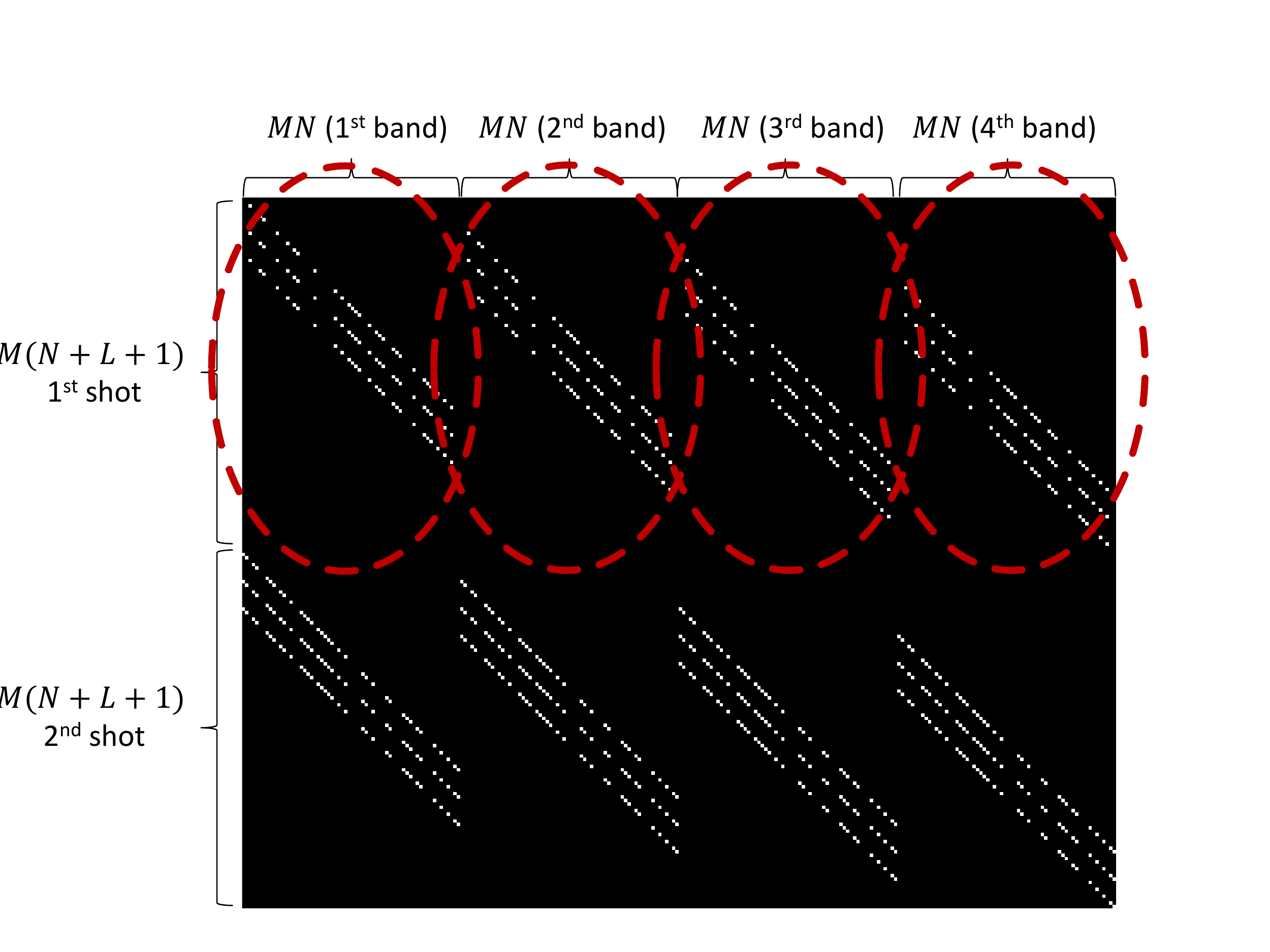}
	\label{fig.1b}
}
\vspace*{0mm}
\caption{\small\sl 
The matrix ${\bf H}$ is presented for $K=2,M=N=8$, and $L=4$. The circled diagonal patterns that repeat horizontally correspond to the coded aperture pattern used in the first FPA shot. The second coded aperture pattern determines the next set of diagonals. In (a) standard CASSI, each FPA shot captures $M(N+L-1)=88$ measurements; in (b) higher order CASSI, each FPA shot captures $M(N+L+1)=104$ measurements.
}
\label{fig.1}
\end{figure}

\subsection{Higher order CASSI}
\label{subsec:HighCASSI}

Recently, Arguello et al.~\cite{Arguello2013higher} proposed a higher order model to characterize the CASSI system with greater precision, and improved the quality of the reconstructed 3D image cubes. In the standard CASSI system model, each cubic voxel in the 3D cube contributes to exactly one measurement in the FPA. In the higher order CASSI model, however, each cubic voxel is shifted to an oblique voxel because of the continuous nature of the 
dispersion, and therefore the oblique voxel contributes to more than one measurement in the FPA. As a result, 
the matrix ${\bf H}$ in~\eqref{eq:CASSI} will have multiple diagonals as shown in Figure~\ref{fig.1b}, where there are sets of $3$ diagonals for each FPA shot, accounting for the voxel energy impinging into the neighboring FPA pixels. 
In this case, the number of measurements with $K=1$ shot of CASSI will be $m=M(N+L+1)$, because each diagonal entails the use of $M$ more pixels 
(we refer readers to~\cite{Arguello2013higher} for details).

In Section~\ref{sec:NumSim}, we will provide promising image reconstruction results for this higher order CASSI system. Using the standard CASSI model, our proposed algorithm produces similar advantageous results over other competing algorithms.

\section{Proposed Algorithm}
\label{sec:Algo}

The goal of our proposed algorithm is to reconstruct the image cube ${\bf f_0}$ from its compressive measurements ${\bf g}$, where the matrix ${\bf H}$ is known.
In this section, we describe our algorithm in detail. The algorithm employs ({\em i}) approximate message passing (AMP)~\cite{DMM2009}, an iterative algorithm for compressive sensing problems, and ({\em ii}) adaptive Wiener filtering, a hyperspectral image denoiser that can be applied within each iteration of AMP.

\subsection{Image denoising in scalar channels}
\label{subsec:scalarChannel}

Below we describe that the linear imaging system model in~\eqref{eq:CASSI} can be converted to a 3D image denoising problem in scalar channels. Therefore, we begin by defining scalar channels, where the noisy observations~${\bf q}$ of the image cube~${\bf f_0}$
obey
\begin{equation}
{\bf q=f_0+v},
\label{eq:scalar}
\end{equation}
and ${\bf v}$ is the additive noise vector. Recovering ${\bf f_0}$ from ~${\bf q}$ is known as a 3D image denoising problem.

\subsection{Approximate message passing}
\label{subsec:AMP}

{\bf Algorithm framework:}
AMP~\cite{DMM2009} has recently become a popular algorithm for solving signal reconstruction problems in linear systems as
defined in~\eqref{eq:CASSI}. 
The AMP algorithm proceeds iteratively according to
\begin{align}
{\bf f}^{t+1}&=\eta_t({\bf H}^T{\bf r}^t+{\bf f}^t)\label{eq:AMPiter1},\\
{\bf r}^t&={\bf g}-{\bf Hf}^t+\frac{1}{R}{\bf r}^{t-1}
\langle\eta_{t-1}'({\bf H}^T{\bf r}^{t-1}+{\bf f}^{t-1})\rangle\label{eq:AMPiter2},
\end{align}
where~${\bf H}^T$ is the transpose of ${\bf H}$, $R=m/n$ represents the measurement rate, $\eta_t(\cdot)$ is a denoising function at the $t$-th iteration, $\eta_t'({\bf s})=\frac{\partial}{\partial {\bf s}}\eta_t({\bf s})$, and~$\langle{\bf u}\rangle=\frac{1}{n}\sum_{i=1}^n u_i$
for some vector~${\bf u}=(u_1,u_2,\ldots,u_n)$.
We will explain in Section~\ref{subsec:deriv} how ${\bf f}^t$ and ${\bf r}^t$ are initialized.
The last term in~\eqref{eq:AMPiter2} is called the ``Onsager reaction term"~\cite{Thouless1977,DMM2009} in statistical physics.
This Onsager reaction term helps improve the phase transition (trade-off between the measurement rate and signal sparsity) of the reconstruction process over existing iterative thresholding algorithms~\cite{DMM2009}.
In the~$t$-th iteration, we obtain the estimated image cube~${\bf f}^t$ and the residual~${\bf r}^t$. 
We highlight that the vector~${\bf H}^T{\bf r}^t+{\bf f}^t$ in~\eqref{eq:AMPiter1} can be regarded as a noise-corrupted version of~${\bf f_0}$ in the~$t$-th iteration with noise variance~$\sigma_t^2$, and therefore~$\eta_t(\cdot)$ is a 3D image denoising function that is performed on a scalar channel as in~\eqref{eq:scalar}. Let us denote the equivalent scalar channel at iteration~$t$ by
\begin{equation}
{\bf q}^t = {\bf H}^T{\bf r}^t +{\bf f}^t= {\bf f_0} + {\bf v}^t,
\label{eq:scalar_t}
\end{equation}
where the noise level $\sigma^2_t$ is estimated by~\cite{Montanari2012},
\begin{equation}
\widehat{\sigma}^2_t=\frac{1}{m}\sum_{i=1}^m (r^t_i)^2,\label{eq:sigma_t}
\end{equation}
and~$r^t_i$ denotes the $i$-th component of the vector~${\bf r}^t$ in~\eqref{eq:AMPiter2}.

{\bf Theoretical properties:}
AMP can be interpreted as minimizing a Gaussian approximation of the Kullback-Leibler divergence~\cite{Cover06} between the estimated and the true posteriors subject to a first order and a second order moment matching constraints between ${\bf f_0}$ and ${\bf Hf_0}$~\cite{Rangan2013ISIT}. If the measurement matrix ${\bf H}$ is i.i.d. Gaussian and the empirical distribution of ${\bf f_0}$ converges to some distribution on $\mathbb{R}$, then the sequence of the mean square error achieved by AMP at each iteration converges to the information theoretical minimum mean square error asymptotically~\cite{Bayati2011}.

Moreover, 
if the matrix is i.i.d. random,
then the noise in the scalar channel~\eqref{eq:scalar} can be viewed as asymptotically i.i.d. Gaussian~\cite{DMM2009,Montanari2012,Krzakala2012probabilistic}.

\subsection{Damping}

We have discussed in Section~\ref{subsec:AMP} that many mathematical properties of AMP hold for the setting where the measurement matrix is i.i.d. Gaussian. When the measurement matrix is not i.i.d. Gaussian, such as the highly structured matrix ${\bf H}$ defined in~\eqref{eq:CASSI}, AMP may encounter divergence issues.
A standard technique called ``damping"~\cite{Rangan2014ISIT,Vila2014} is frequently employed to solve for the divergence problems of AMP, because it only increases the runtime modestly.

Specifically, damping is an extra step within AMP iterations. In~\eqref{eq:AMPiter1}, instead of updating the value of~${\bf f}^{t+1}$ by the output of the denoiser~$\eta_t({\bf H}^T{\bf r}^{t}+{\bf f}^{t})$, we assign a weighted average of~$\eta_t({\bf H}^T{\bf r}^{t}+{\bf f}^{t})$ and~${\bf f}^t$ to~${\bf f}^{t+1}$ as follows,
\begin{equation}
{\bf f}^{t+1} = \alpha\cdot\eta_t({\bf H}^T{\bf r}^{t}+{\bf f}^{t})+(1-\alpha)\cdot{\bf f}^t,
\label{eq:damping1}
\end{equation}
for some constant~$0<\alpha\le1$. Similarly, after obtaining~${\bf r}^t$ in~\eqref{eq:AMPiter2}, we add an extra damping step that updates the value of ${\bf r}^t$ to be $\alpha\cdot {\bf r}^t + (1-\alpha) \cdot {\bf r}^{t-1}$, where the value of $\alpha$ is the same as that in~\eqref{eq:damping1}.

AMP has been proved~\cite{Rangan2014ISIT} to converge with sufficient damping, 
under the assumption that the prior of ${\bf f_0}$ is i.i.d. Gaussian with fixed means and variances throughout all iterations,
and the amount of damping depends on the condition number of the matrix ${\bf H}$. 
Note that other AMP variants~\cite{Swamp2014,Vila2014,RanganADMMGAMP2015} have also been proposed in order to encourage convergence for a broader class of measurement matrices.

\subsection{Adaptive Wiener filter}
\label{subsec:Wiener}

We are now ready to describe our 3D image denoiser, which is the function~$\eta_t(\cdot)$ in the first step of AMP iterations in~\eqref{eq:AMPiter1}. 

{\bf Sparsifying transform:}
Recall that in 2D image denoising problems, a 2D wavelet transform is often performed, and some shrinkage function is applied to the wavelet coefficients in order to suppress noise~\cite{Donoho1994,Figueiredo2001}. The wavelet transform based image denoising method is effective, because natural images are usually sparse in the wavelet transform domain, i.e., there are only a few large wavelet coefficients and the rest of the coefficients are small. Therefore, large wavelet coefficients are likely to contain information about the image, whereas small coefficients are usually comprised mostly of noise, and so it is effective to denoise by shrinking the small coefficients toward zero and suppressing the large coefficients according to the noise variance. Similarly, in hyperspectral image denoising, we want to find a sparsifying transform such that hyperspectral images have only a few large coefficients in this transform domain. 
Inspired by Arguello and Arce~\cite{Arguello2014},
we apply a wavelet transform to each of the 2D images in a 3D cube, and then apply a discrete cosine transform (DCT) along the spectral dimension, because the 2D slices from different wavelengths are highly correlated. That is, the sparsifying transform ${\bf\Psi}$ can be expressed as a Kronecker product of a DCT transform ${\bf \Phi}$ and a 2D wavelet transform ${\bf W}$, i.e., ${\bf\Psi=\Phi\otimes W}$, and it can be shown that ${\bf\Psi}$ is an orthonormal transform. 
Let ${\bf\theta}_{\bf q}^t$ denote the coefficients of ${\bf q}^t$ in this transform domain, i.e., ${\bf\theta}_{{\bf q}}^t={\bf\Psi q}^t$.
Our 3D image denoising procedure will be applied to the coefficients ${\bf\theta}_{\bf q}^t$.
Besides 2D wavelet transform and 1D DCT, it is also possible to sparsify 3D image cubes by dictionary learning~\cite{Yuan2015Side} or Gaussian mixture models~\cite{Rajwade2013}. Moreover, using an endmember mixing matrix~\cite{Martin2015} is an alternative to DCT for characterizing the spectral correlation of 3D image cubes. In this work, we focus on a 2D wavelet transform and 1D DCT as the sparsifying transform, because it is an efficient transform that does not depend on any particular types of image cubes, and an orthonormal transform that is suitable for the AMP framework.

{\bf Parameter estimation in the Wiener filter:}
In our previous work~\cite{Tan_CompressiveImage2014} on compressive imaging reconstruction problems for 2D images,
one of the image denoisers we employed was an adaptive Wiener filter in the wavelet domain, where the variance of each wavelet coefficient was estimated from its neighboring coefficients within a $5\times 5$ window, i.e., the variance was estimated locally. 

As an initial attempt, we applied the previously proposed AMP-Wiener to the reconstruction problem in the CASSI system defined in~\eqref{eq:CASSI}. More specifically, the previously proposed adaptive Wiener filter is applied to the noisy coefficients $\theta_{\bf q}^t$.
Unfortunately, AMP-Wiener encounters divergence issues for the CASSI system  even with significant damping such as $\alpha=0.01$ in~\eqref{eq:damping1}. AMP-Wiener diverges, because it is designed for the setting where the measurement matrix is i.i.d. Gaussian, whereas the measurement matrix ${\bf H}$ defined in~\eqref{eq:CASSI} is highly structured and not i.i.d., and we found in our numerical experiments that the scalar channel noise ${\bf v}^t$ in~\eqref{eq:scalar_t} is not i.i.d. Gaussian. On the other hand, because the Wiener filter allows to conveniently calculate the Onsager term in~\eqref{eq:AMPiter2}, we are motivated to keep the Wiener filter strategy, although the scalar channel \eqref{eq:scalar_t} does not contain i.i.d. Gaussian noise.
Seeing that estimating the coefficient variance from its neighboring coefficients (a $3\times 3$ or $5\times 5$ neighboring window) does not produce reasonable reconstruction for the CASSI system, we modify the local variance estimation to a global estimation within each wavelet subband.
The coefficients $\widehat{\bf\theta}_{\bf f}^t$ of the estimated (denoised) image cube ${\bf f}^t$ are obtained by Wiener filtering,
which can be interpreted as the conditional expectation of ${\bf\theta}_{\bf f}$ given ${\bf\theta}_{\bf q}^t$ under the assumption of Gaussian prior and Gaussian noise,
\begin{eqnarray}
\widehat{\bf\theta}_{{\bf f},i}^t
&=&\frac{\max\{0,\widehat{\nu}_{i,t}^2-\widehat{\sigma}_t^2\}}{(\widehat{\nu}_{i,t}^2-\widehat{\sigma}_t^2)+\widehat{\sigma}_t^2}\left(\theta_{{\bf q},i}^t-\widehat{\mu}_{i,t}\right)+\widehat{\mu}_{i,t}\nonumber\\
&=&\frac{\max\{0,\widehat{\nu}_{i,t}^2-\widehat{\sigma}_t^2\}}{\widehat{\nu}_{i,t}^2}\left(\theta_{{\bf q},i}^t-\widehat{\mu}_{i,t}\right)+\widehat{\mu}_{i,t},
\label{eq:Wiener}
\end{eqnarray}
where ${\bf\theta}_{{\bf q},i}^t$ is the $i$-th element of~${\bf\theta}_{\bf q}^t$, and $\widehat{\mu}_{i,t}$ and $\widehat{\nu}_{i,t}^2$ are the empirical mean and variance of ${\bf \theta}_{{\bf q},i}^t$ within an appropriate wavelet subband, respectively. Taking the maximum between 0 and $(\widehat{\nu}_{i,t}^2-\widehat{\sigma}_t^2)$ ensures that if the empirical variance $\widehat{\nu}_{i,t}^2$ of the noisy coefficients is smaller than the noise variance $\widehat{\sigma}_t^2$, then the corresponding noisy coefficients are set to 0. After obtaining the denoised coefficients~$\widehat{\bf\theta}_{\bf f}^t$, the estimated image cube in the $t$-th iteration satisfies ${\bf f}^t={\bf \Psi}^{-1}\widehat{\bf\theta}_{\bf f}^t={\bf\Psi}^T\widehat{\bf\theta}_{\bf f}^t$. Therefore, the adaptive Wiener filter as a denoiser function~$\eta_t(\cdot)$ can be written as
\begin{eqnarray}
{\bf f}^{t+1} &=&\eta_t({\bf q}^t)\nonumber\\
&=& \boldsymbol\Psi^T\left(\max\{{\bf 0},\widehat{\bf V}_t - \widehat{\sigma}_t^2{\bf I}\}\widehat{\bf V}_t^{-1}
\left(\boldsymbol\Psi{\bf q}^t - \widehat{\boldsymbol\mu}_t\right) + \widehat{\boldsymbol\mu}_t\right),\nonumber\\
\label{eq.etaWiener}
\end{eqnarray}
where {\bf 0} is a zero matrix, $\widehat{\bf V}_t$ is a diagonal matrix with $\widehat{\nu}_{i,t}^2$ on its diagonal, ${\bf I}$ is the identify matrix, $\widehat{\boldsymbol\mu}_t$ is a vector that contains $\widehat{\mu}_{i,t}$, and $\max\{\cdot,\cdot \}$ is operating entry-wise.

We apply this modified adaptive Wiener filter within AMP, and call the algorithm ``AMP-3D-Wiener." We will show in Section~\ref{sec:NumSim} that only a moderate amount of damping is needed for AMP-3D-Wiener to converge.

\subsection{Derivative of adaptive Wiener filter}
\label{subsec:deriv}

The adaptive Wiener filter described in Section~\ref{subsec:Wiener} is applied in~\eqref{eq:AMPiter1} as the 3D image denoising function $\eta_t(\cdot)$. The following step in~\eqref{eq:AMPiter2} requires~$\eta'_t(\cdot)$, i.e., the derivative of~$\eta_t(\cdot)$. We now show how to obtain~$\eta'_t(\cdot)$.
It has been discussed~\cite{Tan_CompressiveImage2014} that when the sparsifying transform is orthonormal, the derivative calculated in the transform domain is equivalent to the derivative in the image domain.
According to~\eqref{eq:Wiener}, the derivative of the Wiener filter in the transform domain with respect to $\widehat{\bf\theta}_{{\bf q},i}^t$ is~$\max\{0,\widehat{\nu}_{i,t}^2-\widehat{\sigma}_t^2\}/\widehat{\nu}_{i,t}^2$.
Because the sparsifying transform ${\bf\Psi}$ is orthonormal, the Onsager term in~\eqref{eq:AMPiter2} can be calculated efficiently 
as
\begin{equation}
\langle\eta'_t({\bf q}^t)\rangle = \frac{1}{n} \sum_{i\in \mathcal{I}}\frac{\max\{0,\widehat{\nu}_{i,t}^2-\widehat{\sigma}_t^2\}}{\widehat{\nu}_{i,t}^2},
\label{eq:Onsager}
\end{equation}
where $\mathcal{I}$ is the index set of all image cube elements, and the cardinality of $\mathcal{I}$ is $n=MNL$.

We focus on image denoising in an orthonormal transform domain and apply Wiener filtering to suppress noise, because it is convenient to obtain the Onsager correction term in~\eqref{eq:AMPiter2}. On the other hand, other denoisers that are not wavelet-DCT based can also be applied within the AMP framework. Metzler et al.~\cite{Metzler2014}, for example, proposed to utilize a block matching and 3D filtering denoising scheme (BM3D)~\cite{Dabov2007} within AMP for 2D compressive imaging reconstruction, and run Monte Carlo~\cite{Ramani2008} to approximate the Onsager correction term. However, the Monte Carlo technique is accurate only when the scalar channel~\eqref{eq:scalar_t} is Gaussian. In the CASSI system model~\eqref{eq:CASSI}, BM4D~\cite{Maggioni2013} may be an option for the 3D image denoising procedure. However, because the matrix~${\bf H}$ is ill-conditioned, the scalar channel~\eqref{eq:scalar_t} that is produced by AMP iterations~(\ref{eq:AMPiter1},\ref{eq:AMPiter2}) is not Gaussian, and thus the Monte Carlo technique fails to approximate the Onsager correction term.

Having completed the description of AMP-3D-Wiener, we summarize AMP-3D-Wiener in Algorithm~\ref{algo:amp_wiener}, where $\widehat{\bf f}_\text{AMP}$ denotes the image cube reconstructed by AMP-3D-Wiener.
Note that in the first iteration of Algorithm~\ref{algo:amp_wiener}, initialization of ${\bf q}^0$ and $\widehat{\sigma}^2_0$ may not be necessary, because ${\bf r}^0$ is an all-zero vector, and the Onsager term is 0 at iteration 1.

\begin{algorithm}[h]
\caption{AMP-3D-Wiener}
\label{algo:amp_wiener}
\textbf{Inputs:} ${\bf g}$, ${\bf H}$, $\alpha$, maxIter\\
{\bf Outputs:} $\widehat{{\bf f}}_\text{AMP}$\\
\textbf{Initialization:} ${\bf f}^1={\bf 0}$, ${\bf r}^{0}={\bf 0}$
\begin{algorithmic}
\For{$t=1:\text{maxIter}$}\\
\begin{enumerate}
\item
${\bf r}^t={\bf g}-{\bf Hf}^t+\frac{1}{R}{\bf r}^{t-1}
\frac{1}{n} \sum_{i=1}^n \frac{\max\{0,\widehat{\nu}_{i,t-1}^2-\widehat{\sigma}_{t-1}^2\}}{\widehat{\nu}_{i,t-1}^2}$
\item
${\bf r}^t = \alpha\cdot{\bf r}^t + (1-\alpha)\cdot{\bf r}^{t-1}$
\item
${{\bf q}}^t={\bf H}^T{\bf r}^t+{\bf f}^t$
\item
$\widehat{\sigma}^2_t=\frac{1}{m}\sum_j ({r}^t_j)^2$
\item
$\theta_{\bf q}^t = {\bf\Psi}{\bf q}^t$
\item
$\widehat{\bf\theta}_{{\bf f},i}^t=\frac{\max\{0,\widehat{\nu}_{i,t}^2-\widehat{\sigma}_t^2\}}{\widehat{\nu}_{i,t}^2}\left(\theta_{{\bf q},i}^t-\widehat{\mu}_{i,t}\right)+\widehat{\mu}_{i,t}$
\item
${\bf f}^{t+1}=\alpha\cdot{\bf\Psi}^T\widehat{\bf\theta}_{\bf f}^t+(1-\alpha) \cdot{\bf f}^{t}$
\end{enumerate}
\EndFor\\
$\widehat{{\bf f}}_\text{AMP}={\bf f}^{\text{maxIter+1}}$

\end{algorithmic}
\end{algorithm}

\begin{figure*}[t]
\setcounter{cnt01}{2}
\vspace*{-5mm}
\hspace*{-10mm}
\includegraphics[width=190mm]{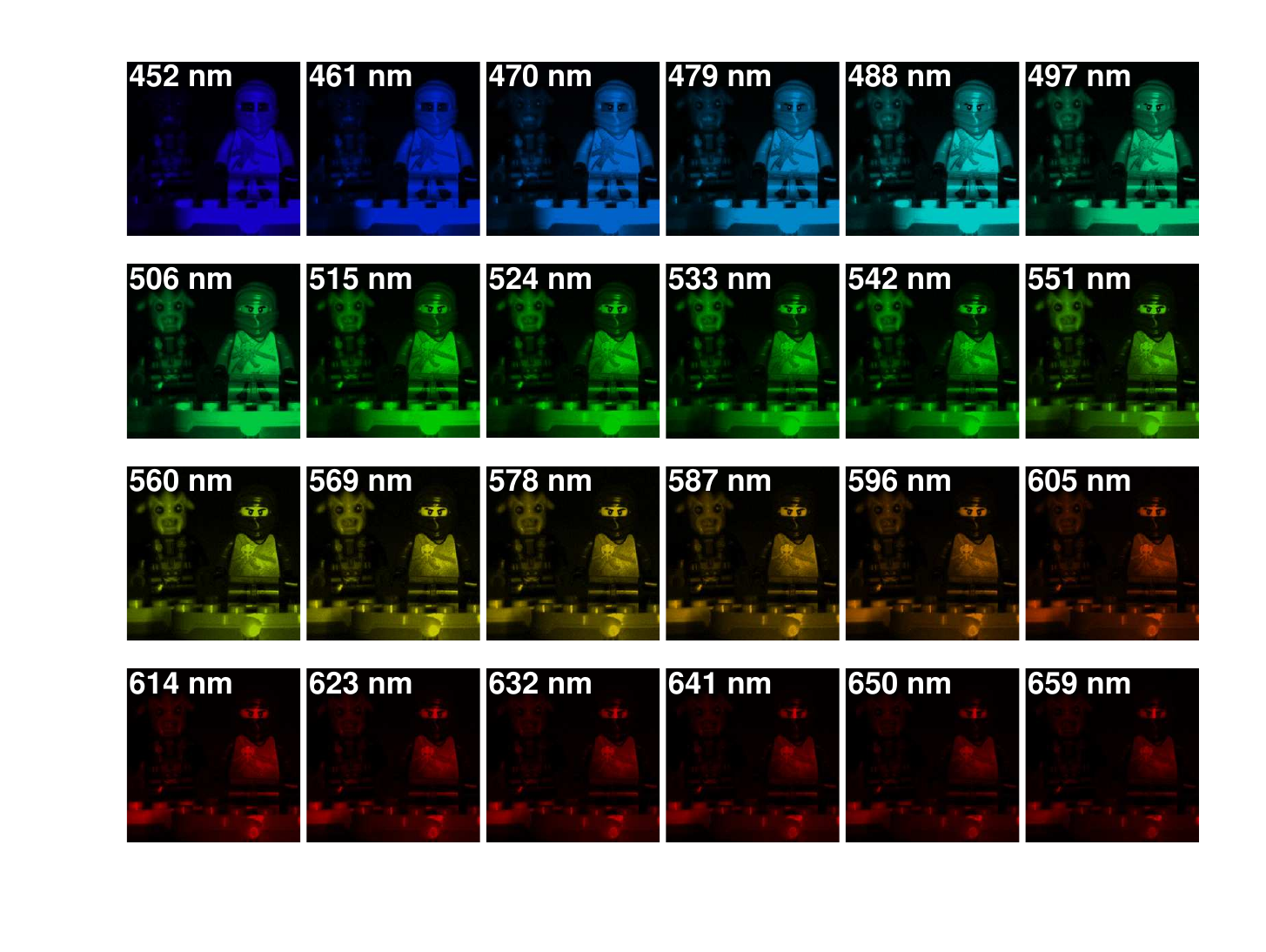}
\vspace*{-15mm}
\caption{\small\sl 
The Lego scene.
(The target object presented in the experimental results was not endorsed by the trademark owners and it is used here as fair use to illustrate the quality of reconstruction of compressive spectral image measurements. LEGO is a trademark of the LEGO Group, which does not sponsor, authorize or endorse the images in this paper. The LEGO Group. All Rights Reserved.
http://aboutus.lego.com/en-us/legal-notice/fair-play/.)
}
\label{fig.Lego}
\setcounter{figure}{\value{cnt01}}
\end{figure*}

\section{Numerical Results}
\label{sec:NumSim}

In this section, we provide numerical results where we compare the reconstruction quality and runtime of AMP-3D-Wiener,
gradient projection for sparse reconstruction (GPSR)~\cite{GPSR2007}, and two-step iterative shrinkage/thresholding (TwIST)~\cite{Wagadarikar2008,NewTWIST2007}.
In all experiments, we use the same coded aperture pattern for AMP-3D-Wiener, GPSR, and TwIST.
In order to quantify the reconstruction quality of each algorithm, the peak signal to noise ratio (PSNR) of each 2D slice in the reconstructed cubes is measured.
The PSNR is defined as the ratio between the maximum squared value of the ground truth image cube~${\bf f_0}$ and the mean square error of the estimation~$\widehat{\bf f}$, i.e.,
\begin{equation*}
\text{PSNR}=10\cdot\log_{10}\left(\frac{\max_{x,y,\lambda}\left(f^2_{0,(x,y,\lambda)}\right)}{\sum_{x,y,\lambda}\left(\widehat{f}_{(x,y,\lambda)}-f_{0,(x,y,\lambda)}\right)^2}\right),
\end{equation*}
where $f_{(x,y,\lambda)}$ denotes the element in the cube ${\bf f}$ at spatial coordinate $(x,y)$ and spectral coordinate $\lambda$.

In AMP, the damping parameter~$\alpha$ is set to be 0.2.
Recall that increasing the amount of damping helps prevent the divergence of AMP-3D-Wiener, and that the divergence issue can be identified by evaluating the values of $\widehat{\sigma}_t^2$ from~\eqref{eq:sigma_t}. We select 0.2 as the damping parameter value, because 0.2 is the maximum damping value such that AMP-3D-Wiener converges in all the image cubes we test.
The divergence issues of AMP-3D-Wiener can be detected by evaluating the value of $\widehat{\sigma}_t^2$ obtained by~\eqref{eq:sigma_t} as a function of iteration number $t$. Recall that $\widehat{\sigma}_t^2$ estimates the amount of noise in the noisy image cube~${\bf q}^t$ at iteration $t$. If AMP-3D-Wiener converges, then we expect the value of $\widehat{\sigma}_t^2$ to decrease as $t$ increases. Otherwise, we know that AMP-3D-Wiener diverges.
The choice of damping mainly depends on the structure of the imaging model in~\eqref{eq:CASSI} but not on the characteristics of the image cubes, and thus the value of the damping parameter~$\alpha$ need not be tuned in our experiments.

To reconstruct the image cube~${\bf f_0}$, GPSR and TwIST minimize objective functions of the form
\begin{equation}
\widehat{\bf f}=\arg\min_{\bf f} \frac{1}{2}\|{\bf g-Hf}\|_2^2 + \beta\cdot\phi({\bf f}),
\label{eq:Obj}
\end{equation}
where $\phi(\cdot)$ is a regularization function that characterizes the structure of the image cube ${\bf f_0}$, and $\beta$ is a regularization parameter that balances the weights of the two terms in the objective function. In GPSR, $\phi({\bf f}) = \|{\bf \Psi f}\|_1$; in TwIST, the total variation regularizer is employed,
\begin{eqnarray}
\phi({\bf f}) &=& \sum_{\lambda=1}^L\sum_{x=1}^M\sum_{y=1}^N
\bigg((f(x+1,y,\lambda)-f(x,y,\lambda))^2 \nonumber\\
&&+ (f(x,y+1,\lambda)-f(x,y,\lambda))^2\bigg)^{1/2}.
\end{eqnarray}
Note that the role of the $\ell_1$-norm of the sparsifying coefficients in GPSR is to impose the overall sparsity of the sparsifying coefficients, whereas the total variation in TwIST encourages spatial smoothness in the reconstructed image cubes.
The implementation of GPSR is downloaded from ``http://www.lx.it.pt/~mtf/GPSR/," and the implementation of TwIST is downloaded from ``http://www.disp.duke.edu/projects/CASSI/experimentaldata/\\
index.ptml."
The value of the regularization parameter~$\beta$ in~\eqref{eq:Obj} greatly affects the reconstruction results of GPSR and TwIST, and must be tuned carefully.
We select the optimal values of $\beta$ for GPSR and TwIST manually, i.e., we run GPSR and TwIST with $5-10$ 
different values of $\beta$, and select the results with the highest PSNR.\footnote{As an example, we simulate GPSR with many different values for $\beta$, and obtain that for $\beta=1\cdot 10^{-5},2\cdot 10^{-5},3\cdot 10^{-5},4\cdot 10^{-5},5\cdot 10^{-5},6\cdot 10^{-5}$, and $7\cdot 10^{-5}$, the corresponding PSNRs of the reconstructed cubes are $31.25$ dB, $32.30$ dB, $32.82$ dB, $32.99$ dB, $33.02$ dB, $33.09$ dB, and $33.06$ dB. Therefore, we select $\beta=6\cdot 10^{-5}$ for this specific image cube. We follow the same procedure to select the optimal $\beta$ values for each test image cube.}
The typical value of the regularization parameter for GPSR is between $10^{-5}$ and $10^{-4}$, and the value for TwIST is around 0.1. 
We note in passing that the ground truth image cube is not known in practice, and estimating the PSNR
obtained using different $\beta$ may be quite involved and require oracle-like information when using GPSR and TwIST.
Reweighted $\ell_1$-minimization~\cite{Candes2008} does not need regularization parameter tuning, and has been shown to outperform $\ell_1$-minimization by Candes et al.~\cite{Candes2008}. However, the existing reweighted $\ell_1$-minimization implementations require either QR decomposition~\cite{Meyer} of the measurement matrix ${\bf H}$ or the null space of ${\bf H}$, 
which requires ${\bf H}$ to be expressed as a matrix. That said, ${\bf H}$ is a very large matrix, and we implement it as a linear operator.
Therefore, implementing the reweighted $\ell_1$-minimization that is applicable to the system model in \eqref{eq:CASSI} is beyond the scope of this paper, and the reweighted $\ell_1$-minimization is not included in our simulation results.
There exist other hyperspectral image reconstruction algorithms based on dictionary learning~\cite{Rajwade2013,Yuan2015Side}.  
In order to learn a dictionary that represents a 3D image, the image cube needs to be divided into small patches, and the measurement matrix~${\bf H}$ also needs to be divided accordingly. Dividing the measurement matrix into smaller patches is convenient for the standard CASSI model (Figure~\ref{fig.1a}), because there is a one-to-one correspondence between the measurement matrix and the image cube, i.e., each measurement is a linear combination of only one voxel in each spectral band. In higher order CASSI, however, each measurement is a linear combination of multiple voxels in each spectral band.
Therefore,
it is not straightforward to modify these dictionary learning methods to the higher order CASSI model described in Section~\ref{subsec:HighCASSI}, and we do not include these algorithms in the comparison.

\subsection{Test on ``Lego" image cube}
\label{subsec:LegoTest}

The first set of simulations is performed for the scene shown in Figure~\ref{fig.Lego}. 
This data cube was acquired using a wide-band Xenon lamp as the illumination source, modulated by a visible monochromator spanning the spectral range between $448$ nm and $664$ nm, and each spectral band has $9$ nm width. The image intensity was captured using a grayscale CCD camera, with pixel size $9.9$ $\mu$m, and 8 bits of intensity levels. The resulting test data cube has $M \times N = 256 \times 256$ pixels of spatial resolution and $L = 24$ spectral bands.

{\bf Setting 1:} The measurements~${\bf g}$ are captured with $K=2$ shots.
The coded aperture in the first shot is generated randomly with 50\% of the aperture being opaque, and
the coded aperture in the second shot is the complement of the aperture in the first shot. 
The measurement rate with two shots is $m/n=KM(N+L+1)/(MNL)\approx0.09$.
Moreover, we add Gaussian noise with zero mean to the measurements. The signal to noise ratio (SNR) is defined as $10\log_{10}(\mu_g/\sigma_\text{noise})$~\cite{Arguello2014}, where $\mu_g$ is the mean value of the measurements ${\bf Hf_0}$ and $\sigma_\text{noise}$ is the standard deviation of the additive noise~${\bf z}$. In Setting 1, we add measurement noise such that the SNR is 20 dB.

We note in passing that 
the complementary random coded apertures are binary, and can be implemented through photomask technology or emulated by a digital micromirror device (DMD). Therefore, the complementary random coded apertures are feasible in practice~\cite{Arguello2014}.
Moreover, 
the complementary random coded apertures ensure that in the matrix ${\bf H}$ in~\eqref{eq:CASSI}, the norm of each column is similar, which is suitable for the AMP framework.
However, it is a limitation of the current AMP-3D-Wiener that the complementary random coded apertures must be employed,
otherwise, AMP-3D-Wiener may diverge.

\begin{figure}[t]
\vspace*{-2mm}
\begin{center}
\includegraphics[width=80mm]{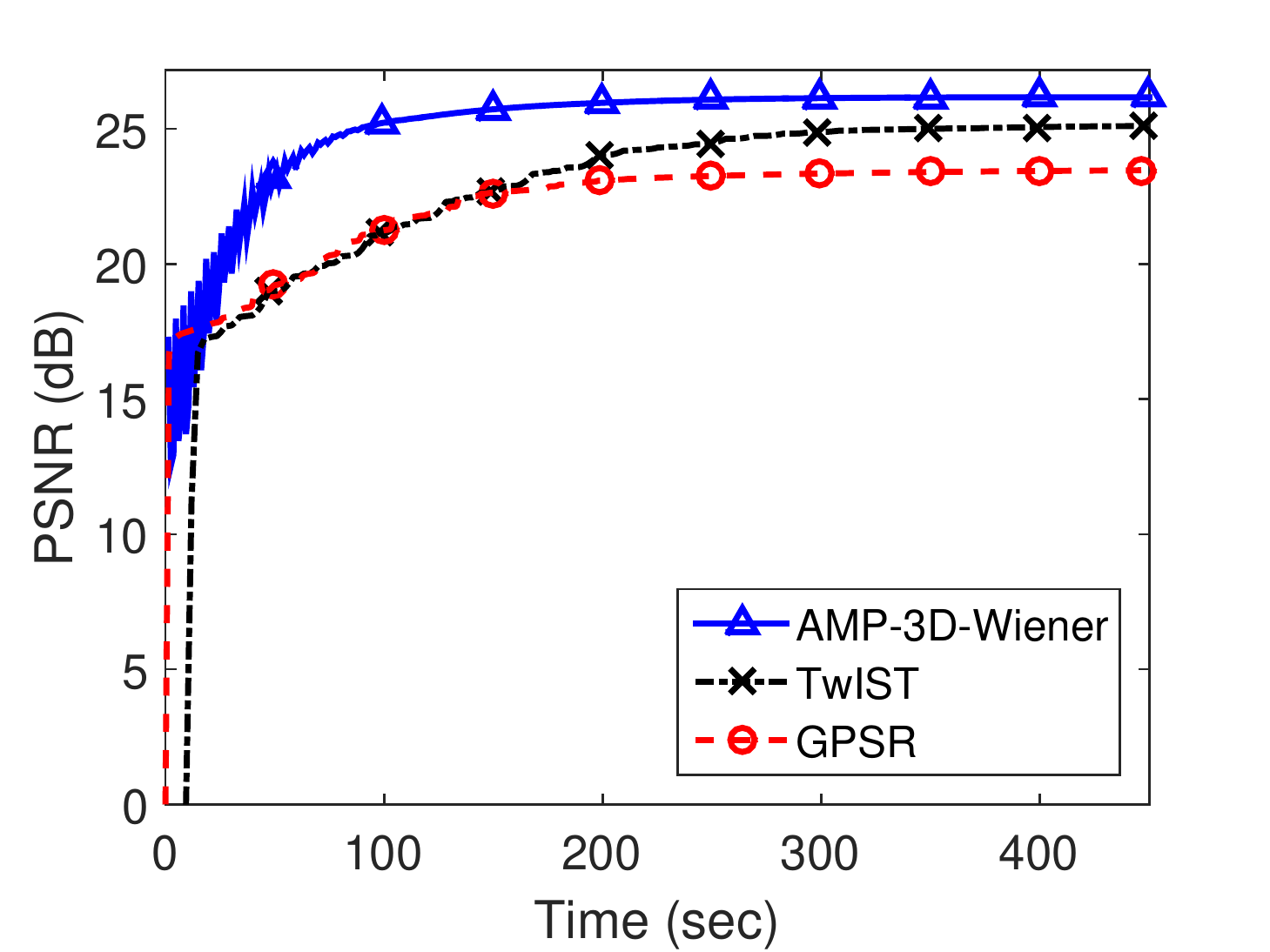}
\end{center}
\vspace*{0mm}
\caption{\small\sl 
Runtime versus average PSNR comparison of AMP-3D-Wiener, GPSR, and TwIST for the Lego image cube. Cube size is $M=N=256$, and $L=24$. The measurements are captured with $K=2$ shots using complementary random coded apertures, and the number of measurements is $m=143,872$. Random Gaussian noise is added to the measurements such that the SNR is 20 dB.}
\label{fig.iter_Psnr}
\end{figure}

Let us now evaluate the numerical results for Setting 1.
Figure~\ref{fig.iter_Psnr} compares the reconstruction quality of AMP-3D-Wiener, GPSR, and TwIST within a certain amount of runtime. 
Runtime is measured on a Dell OPTIPLEX 9010 running an Intel(R)
CoreTM i7-860 with 16GB RAM, and the environment is Matlab R2013a.
In Figure~\ref{fig.iter_Psnr},
the horizontal axis represents runtime in seconds, and the vertical axis is the averaged PSNR over the 24 spectral bands.
Although the PSNR of AMP-3D-Wiener oscillates during the first few iterations, which may be because the matrix~${\bf H}$ is ill-conditioned, it becomes stable after 50 seconds and reaches a higher level when compared to the PSNRs of GPSR and TwIST at 50 seconds.
After 450 seconds, the average PSNR of the cube reconstructed by AMP-3D-Wiener (solid curve with triangle markers) is 26.16 dB, while the average PSNRs of GPSR  (dash curve with circle markers) and TwIST (dash-dotted curve with cross markers) are 23.46 dB and 25.10 dB, respectively. Note that in 450 seconds, TwIST runs roughly 200 iterations, while AMP-3D-Wiener and GPSR run 400 iterations.

Figure~\ref{fig.spectral_Psnr} complements Figure~\ref{fig.iter_Psnr} by illustrating the PSNR of each 2D slice in the reconstructed cube separately. It is shown that the cube reconstructed by AMP-3D-Wiener has $2-4$ dB higher PSNR than the cubes reconstructed by GPSR and $0.4-3$ dB higher than those of TwIST for all 24 slices.

\begin{figure}[t]
\vspace*{-2mm}
\begin{center}
\includegraphics[width=80mm]{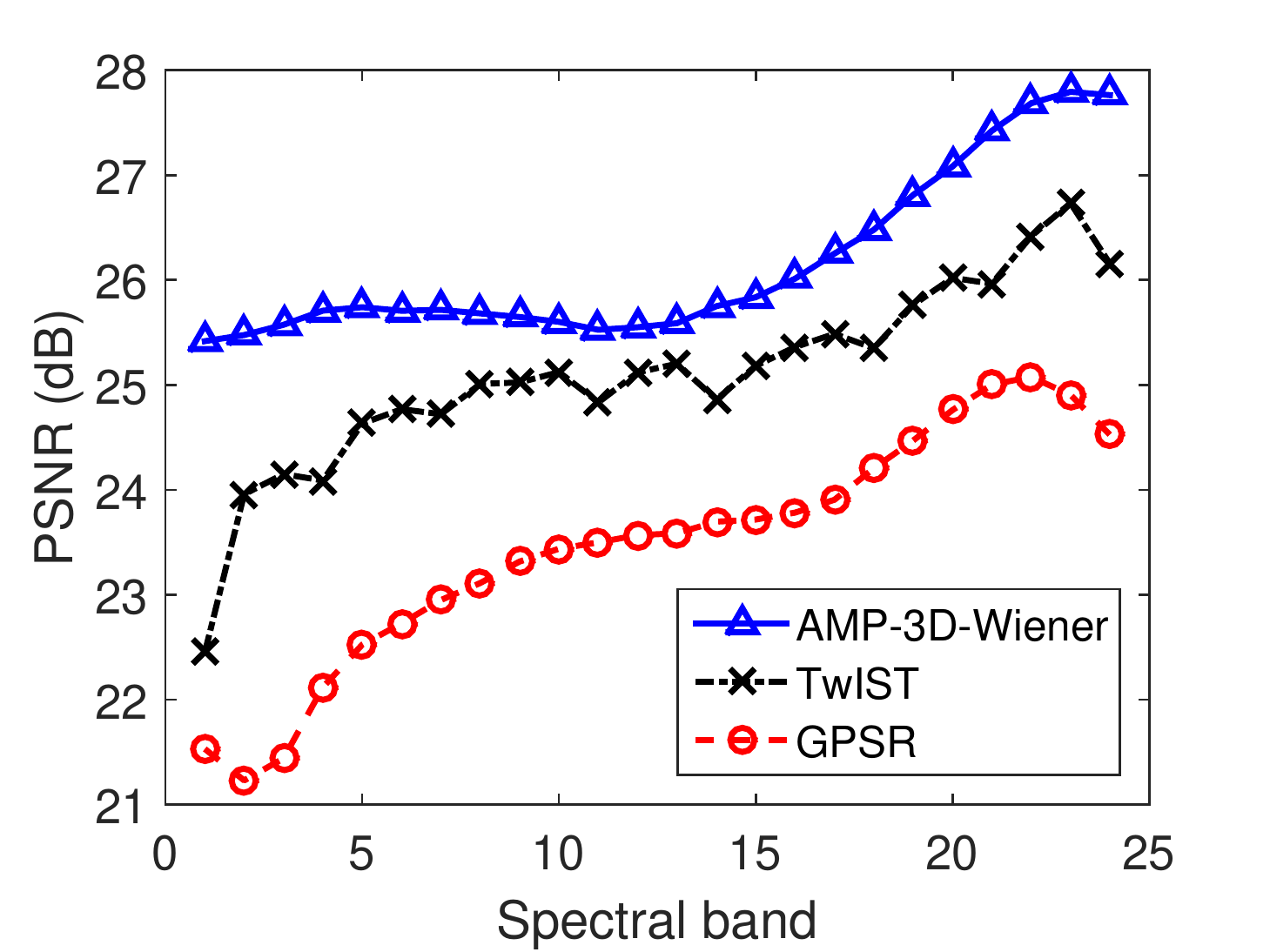}
\end{center}
\vspace*{0mm}
\caption{\small\sl 
Spectral band versus PSNR comparison of AMP-3D-Wiener, GPSR, and TwIST for the Lego image cube. Cube size is $M=N=256$, and $L=24$. The measurements are captured with $K=2$ shots using complementary random coded apertures, and the number of measurements is $m=143,872$. Random Gaussian noise is added to the measurements such that the SNR is 20 dB.}
\label{fig.spectral_Psnr}
\end{figure}

In Figure~\ref{fig.ImgComp}, we plot the 2D slices at wavelengths $488$ nm, $533$ nm, and $578$ nm in the actual image cubes reconstructed by AMP-3D-Wiener, GPSR, and TwIST. The images in these four rows are slices from the ground truth image cube~${\bf f_0}$, the cubes reconstructed by AMP-3D-Wiener, GPSR, and TwIST, respectively.
The images in columns $1-3$ show the upper-left part of the scene, whereas images in columns $4-6$ show the upper-right part of the scene. All images are of size $128\times128$.
It is clear from Figure~\ref{fig.ImgComp} that the 2D slices reconstructed by AMP-3D-Wiener have better visual quality; the slices reconstructed by GPSR have blurry edges, and the slices reconstructed by TwIST lack details, because the total variation regularization tends to constrain the images to be piecewise constant.

\begin{figure*}[t!]
\setcounter{cnt02}{5}
\vspace*{-5mm}
\hspace*{-10mm}
\includegraphics[width=190mm]{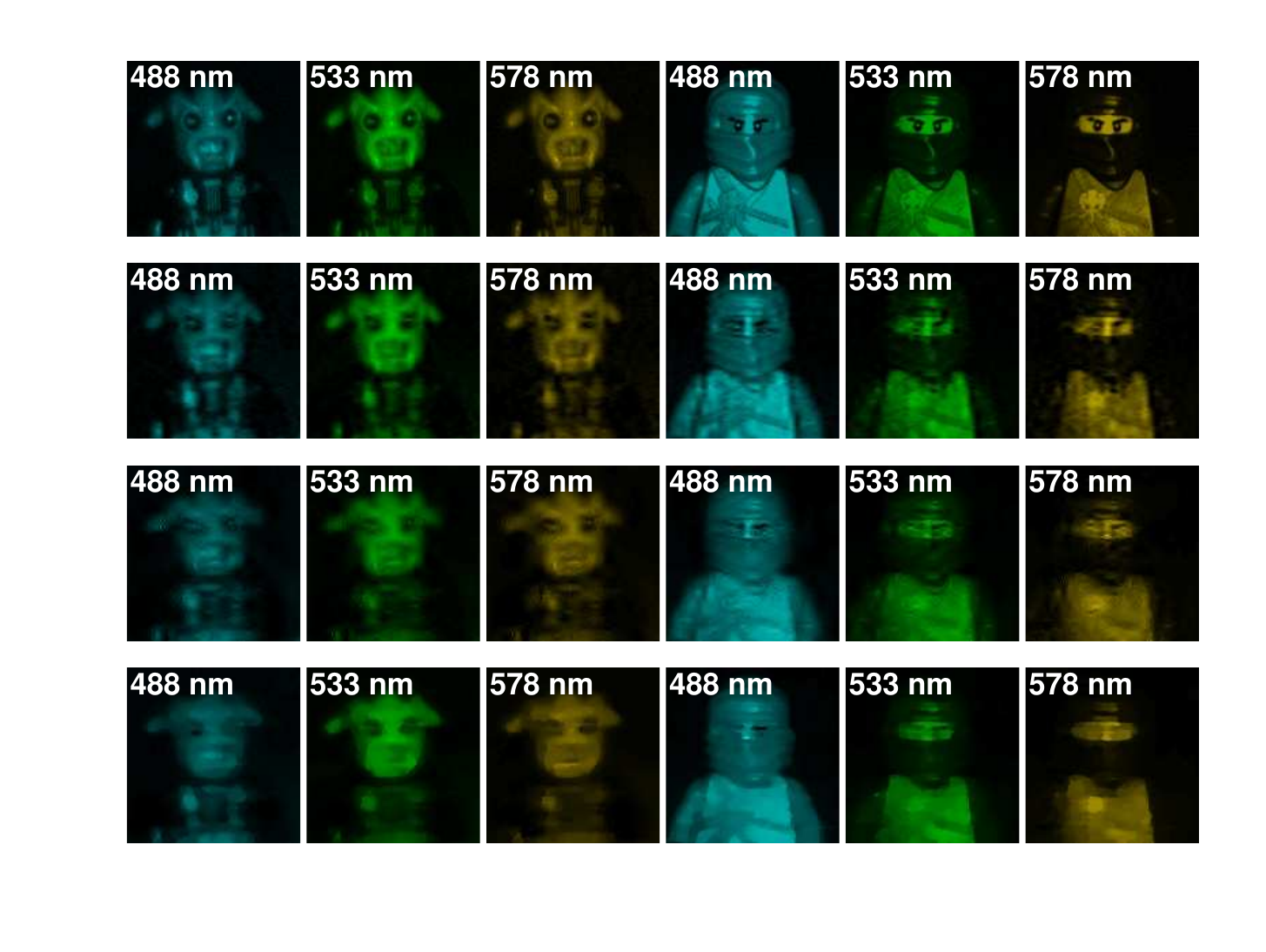}
\vspace*{-15mm}
\caption{\small\sl 
2D slices at wavelengths $488$ nm, $533$ nm, and $578$ nm in the image cubes reconstructed by AMP-3D-Wiener, GPSR, and TwIST for the Lego image cube. Cube size is $M=N=256$, and $L=24$. The measurements are captured with $K=2$ shots using complementary random coded apertures, and the number of measurements is $m=143,872$. Random Gaussian noise is added to the measurements such that the SNR is 20 dB. 
First row: ground truth; second row: the reconstruction result by AMP-3D-Wiener; third row: the reconstruction result by GPSR; last row: the reconstruction result by TwIST.
Columns $1-3$: upper-left part of the scene of size $128\times 128$; columns $4-6$: upper-right part of the scene of size $128\times128$.
}
\label{fig.ImgComp}
\setcounter{figure}{\value{cnt02}}
\end{figure*}

\begin{figure*}[t!]
\setcounter{cnt03}{6}
\vspace*{-0mm}
\hspace*{10mm}
\subfigure[Original image]{
	\includegraphics[width=0.4\textwidth]{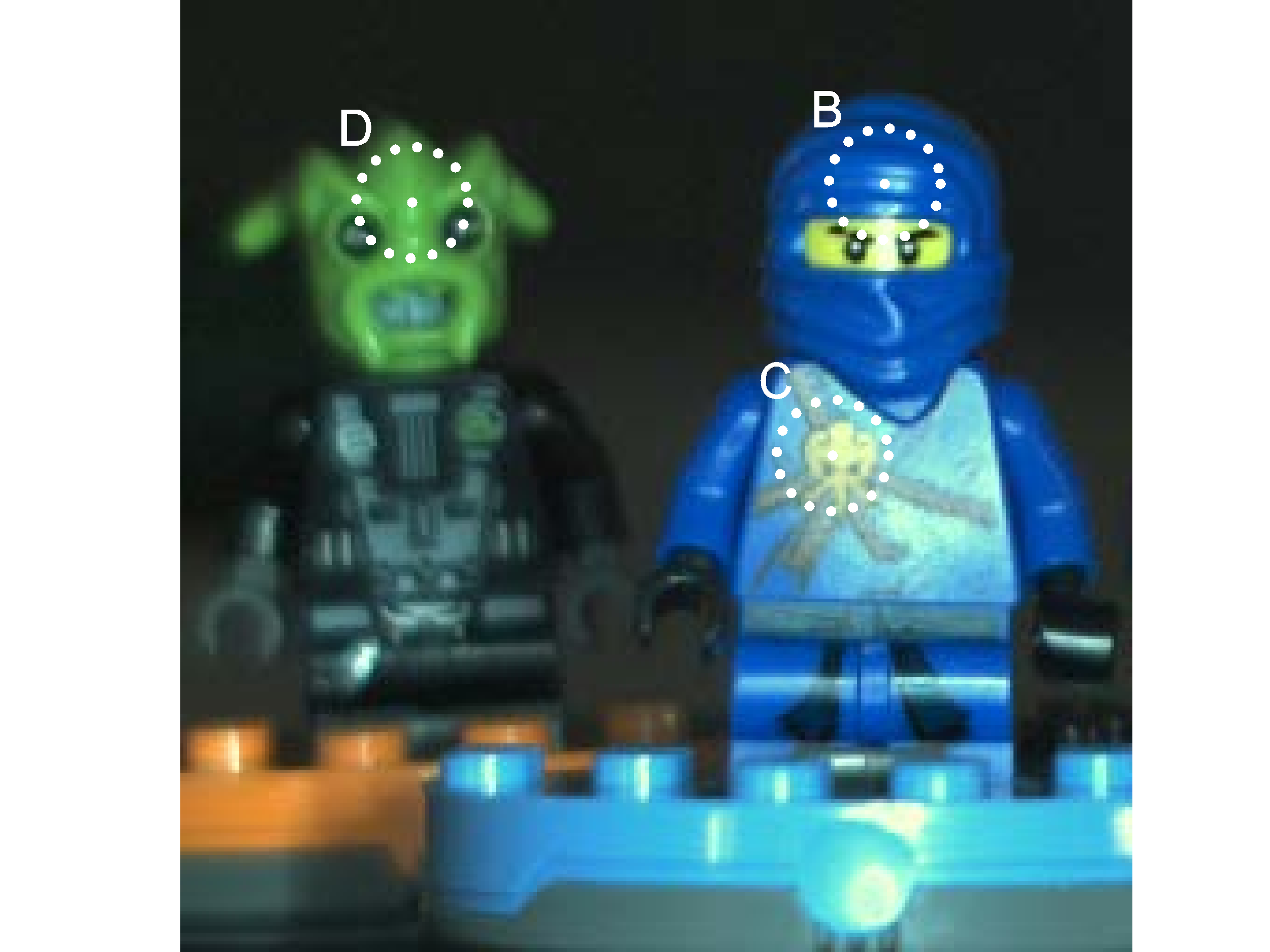}
	\label{fig.BCD}
}
\subfigure[$x=190,y=50$]{
	\includegraphics[width=0.4\textwidth]{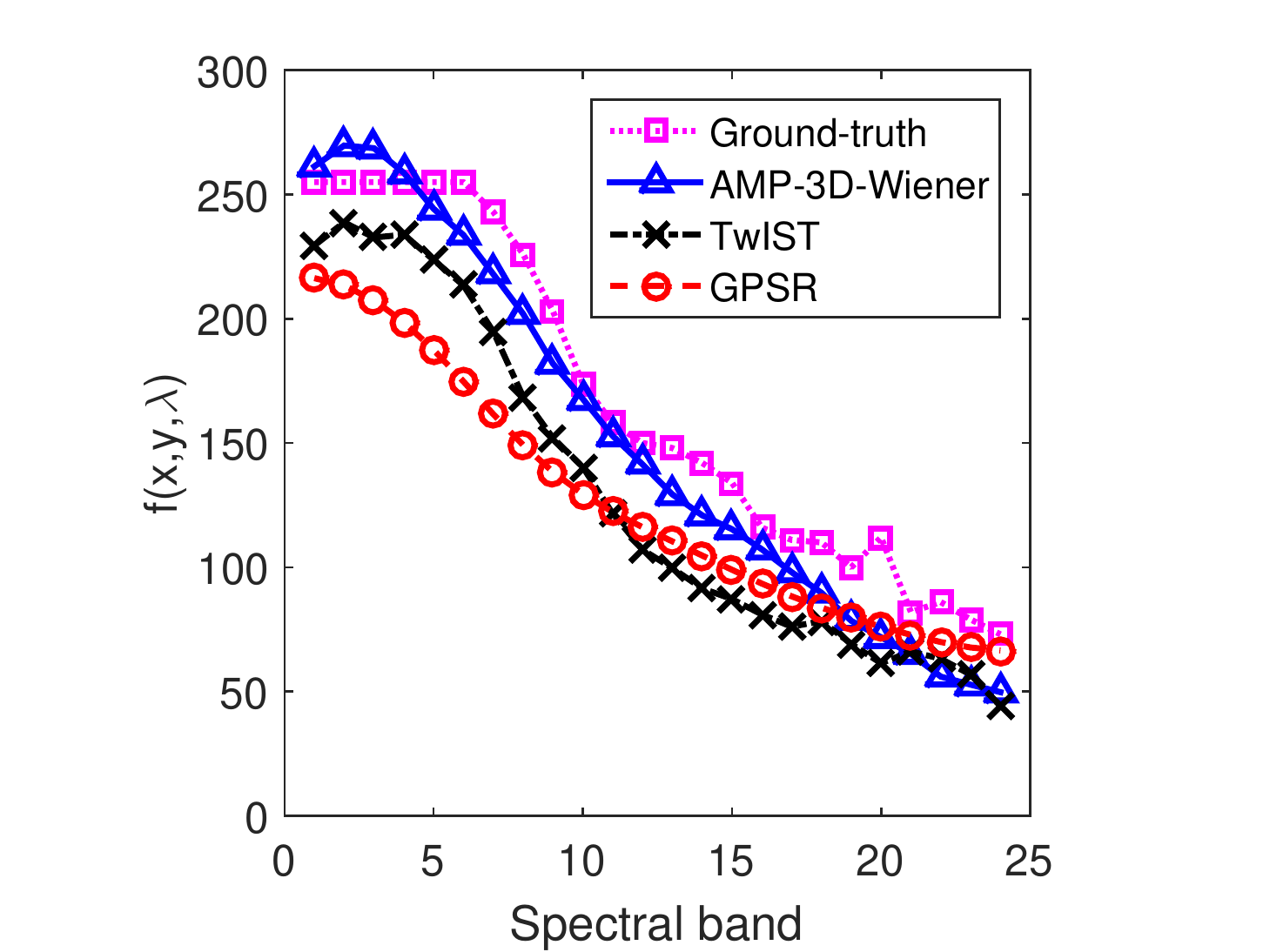}
	\label{fig.B}
}\\
\hspace*{10mm}
\subfigure[$x=176,y=123$]{
	\includegraphics[width=0.4\textwidth]{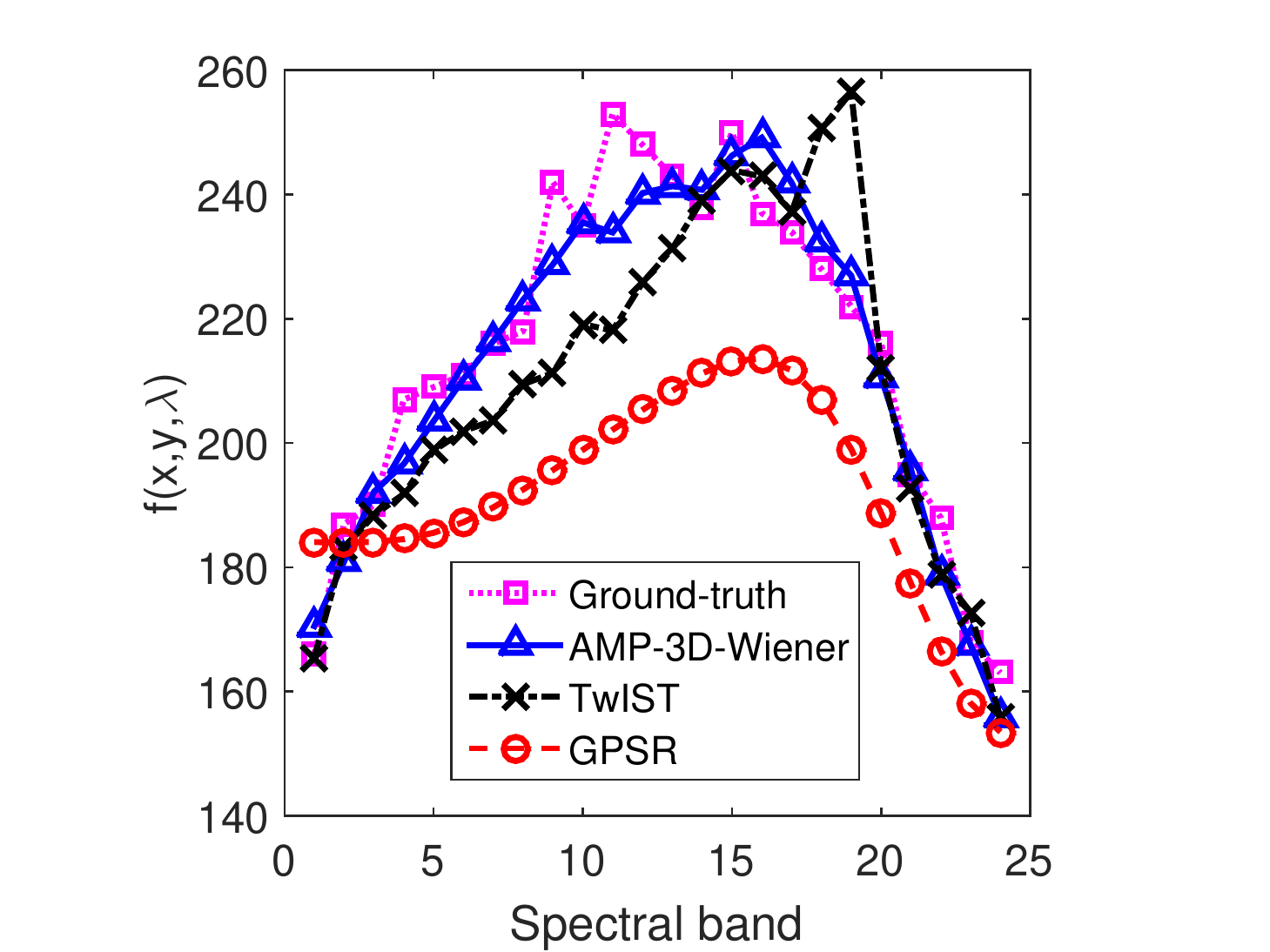}
	\label{fig.C}
}
\subfigure[$x=63,y=55$]{
	\includegraphics[width=0.4\textwidth]{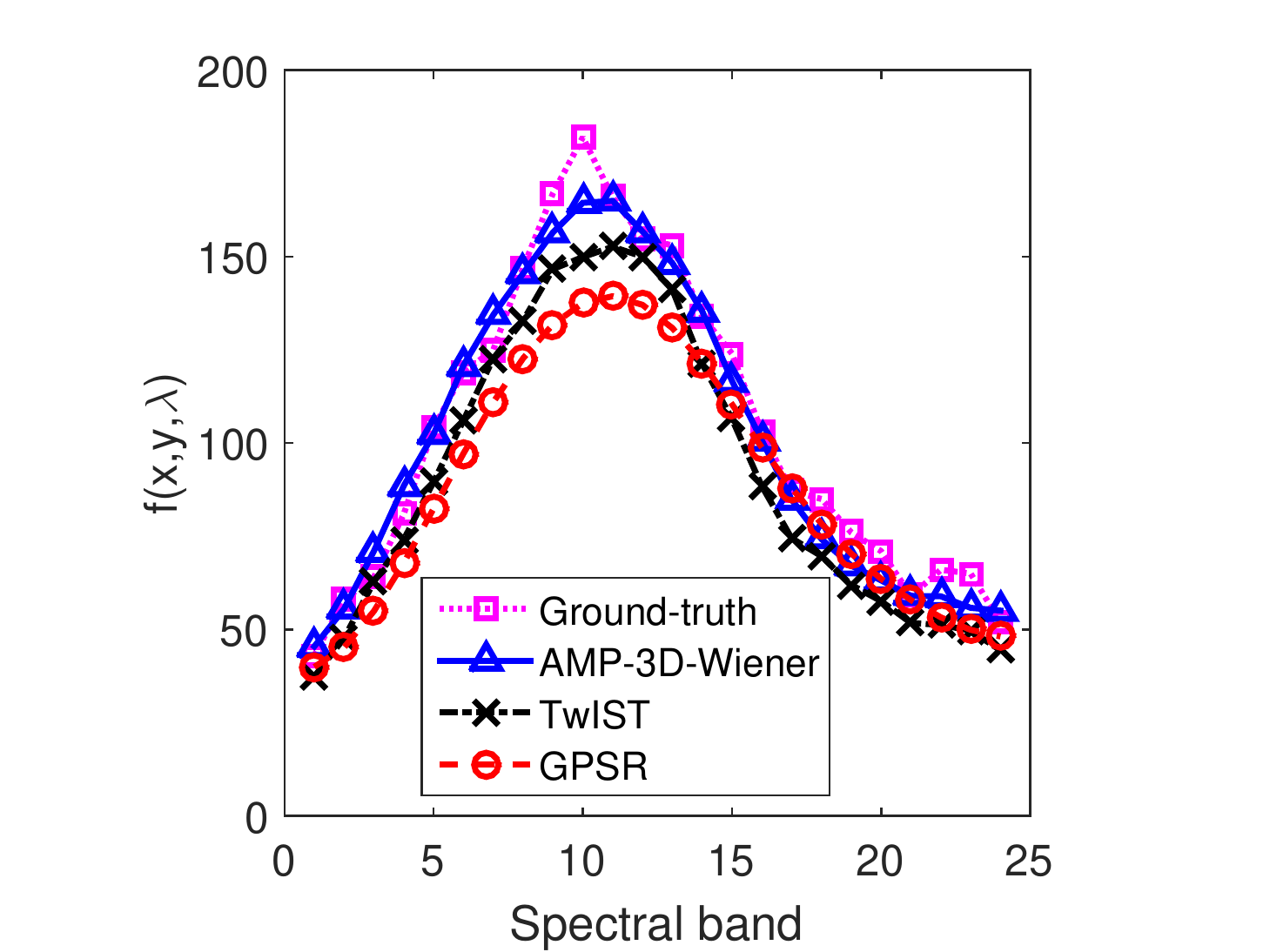}
	\label{fig.D}
}
\caption{\small\sl
Comparison of AMP-3D-Wiener, GPSR, and TwIST on reconstruction along the spectral dimension of three spatial pixel locations as indicated in (a). The estimated pixel values are illustrated for (b) the pixel B, (c) the pixel C, and (d) the pixel D. }
\label{fig.sig}
\setcounter{figure}{\value{cnt03}}
\end{figure*}

Furthermore, a spectral signature plot analyzes how the pixel values change along the spectral dimension at a fixed spatial location, and we present such spectral signature plots for the image cubes reconstructed by AMP-3D-Wiener, GPSR, and TwIST in Figure~\ref{fig.sig}.
Three spatial locations are selected as shown in Figure~\ref{fig.BCD}, and the spectral signature plots for locations B, C, and D are shown in Figures~\ref{fig.B}--\ref{fig.D}, respectively. It can be seen that the spectral signatures of the cube reconstructed by AMP-3D-Wiener closely resemble those of the ground truth image cube (dotted curve with square markers), whereas there are discrepancies between the spectral signatures of the cube reconstructed by GPSR or TwIST and those of the ground truth cube.

According to the runtime experiment from Setting 1, we run AMP-3D-Wiener with 400 iterations, GPSR with 400 iterations, and TwIST with 200 iterations for the rest of the simulations, so that all algorithms complete within a similar amount of time.

\begin{figure}[t]
\vspace*{-5mm}
\begin{center}
\includegraphics[width=80mm]{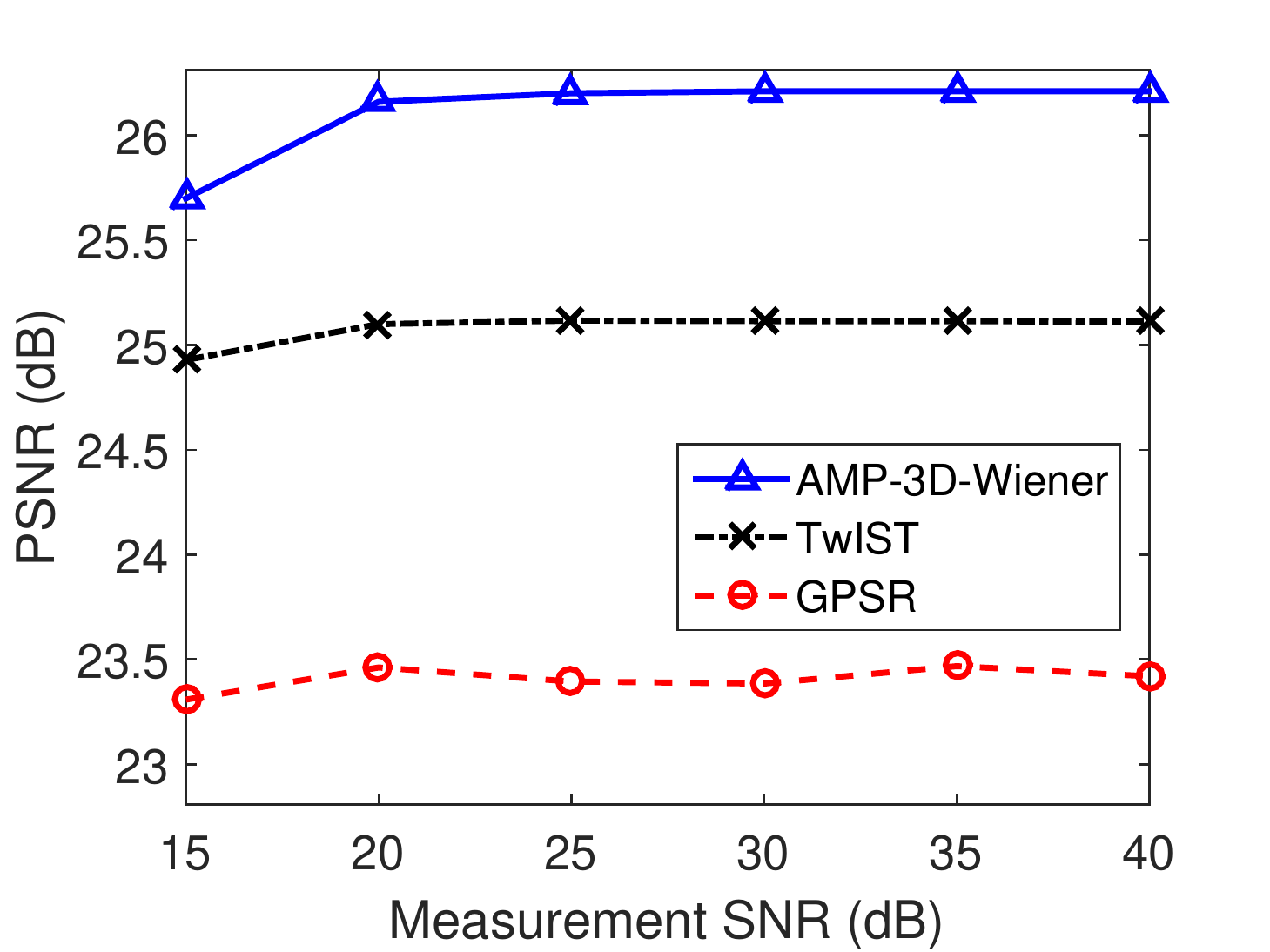}
\end{center}
\vspace*{0mm}
\caption{\small\sl 
Measurement noise versus average PSNR comparison of AMP-3D-Wiener, GPSR, and TwIST for the Lego image cube. Cube size is $M=N=256$, and $L=24$. The measurements are captured with $K=2$ shots using complementary random coded apertures, and the number of measurements is $m=143,872$.}
\label{fig.SNR_Psnr}
\end{figure}

{\bf Setting 2:} In this experiment, we add measurement noise such that the SNR varies from 15 dB to 40 dB, which is the same setting as in Arguello and Arce~\cite{Arguello2014}, and the result is shown in Figure~\ref{fig.SNR_Psnr}. Again, AMP-3D-Wiener achieves more than 2 dB higher PSNR than GPSR, and about 1 dB higher PSNR than TwIST, overall.

{\bf Setting 3:} In Settings 1 and 2, the measurements are captured with $K=2$ shots. We now test our algorithm on the setting where the number of shots varies from $K=2$ to $K=12$ with pairwise complementary random coded apertures. 
Specifically, we randomly generate the coded aperture for the $k$-th shot for $k=1,3,5,7,9,11$, and the coded aperture in the $(k+1)$-th shot is the complement of the aperture in the $k$-th shot.
In this setting, a moderate amount of noise (20 dB) is added to the measurements. Figure~\ref{fig.shots_Psnr} presents the PSNR of the reconstructed cubes as a function of the number of shots, and AMP-3D-Wiener consistently beats GPSR and TwIST.

\begin{figure}[h]
\vspace*{-5mm}
\begin{center}
\includegraphics[width=80mm]{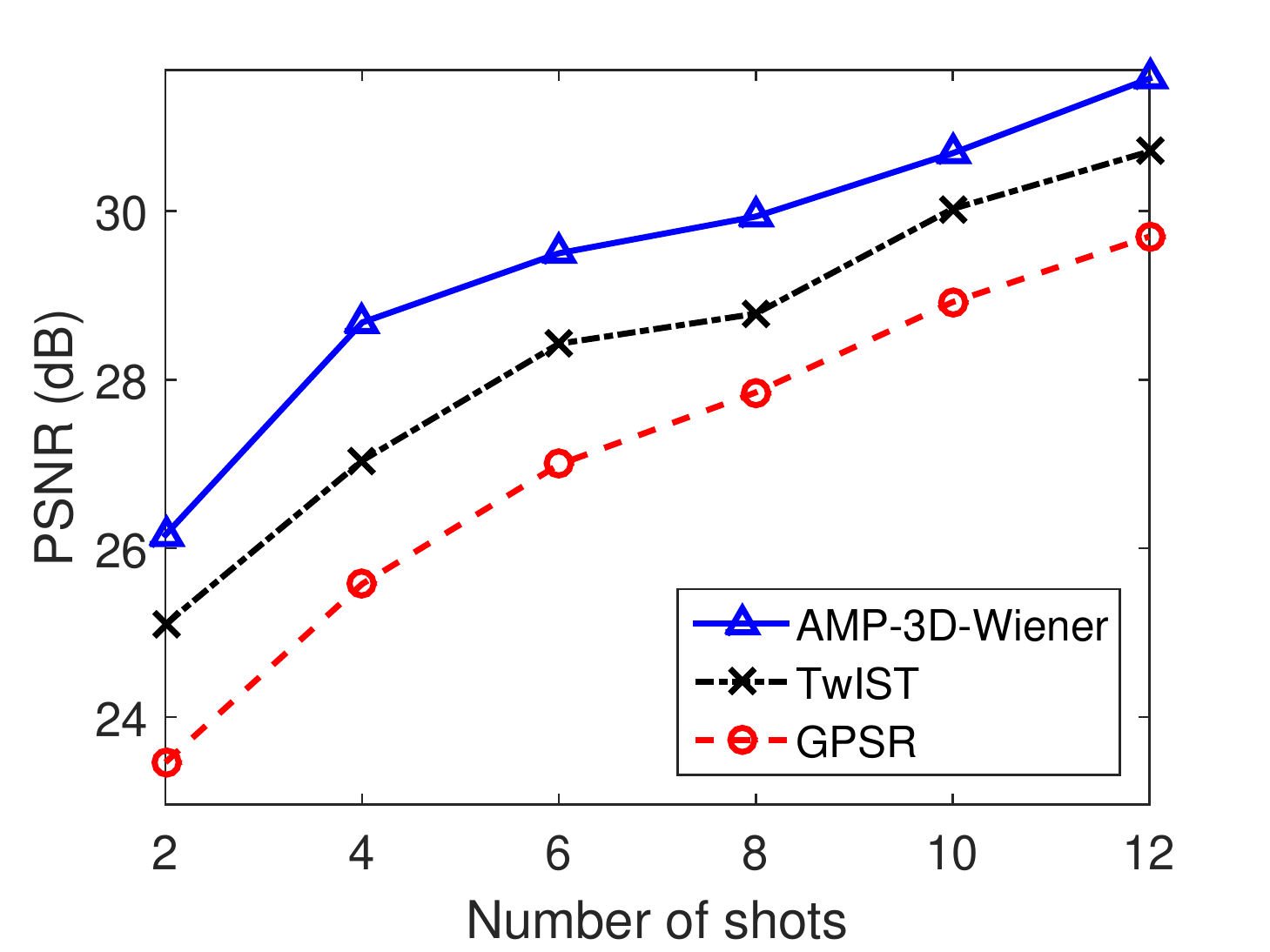}
\end{center}
\vspace*{0mm}
\caption{\small\sl 
Number of shots versus average PSNR comparison of AMP-3D-Wiener, GPSR, and TwIST for the Lego image cube. Cube size is $N=M=256$, and $L=24$. The measurements are captured using pairwise complementary random coded apertures. Random Gaussian noise is added to the measurements such that the SNR is 20 dB.}
\label{fig.shots_Psnr}
\end{figure}

\subsection{Test on natural scenes}
\label{subsec:NaturalTest}

Besides the Lego image cube, we have also tested our algorithm on image cubes of natural scenes~\cite{Foster2006}.\footnote{The cubes are downloaded from http://personalpages.manchester.ac.uk/staff/\\
d.h.foster/Hyperspectral$\_$images$\_$of$\_$natural$\_$scenes$\_$04.html and
http://per-\\sonal
pages.manchester.ac.uk/staff/d.h.foster/Hyperspectral$\_$images$\_$of$\_$natural$\_$\\scenes$\_$02.html.}
There are two datasets, ``natural scenes 2002" and ``natural scenes 2004," 
each one with 8 image data cubes. The cubes in
the first dataset have $L=31$ spectral bands with spatial resolution of around $700\times700$, whereas the cubes in
the second dataset have $L=33$ spectral bands with spatial resolution of around $1000\times 1000$.
To satisfy the dyadic constraint of the 2D wavelet, we crop their spatial resolution to be $M=N=512$.
Because the spatial dimensions of the cubes ``scene 6" and ``scene7" in the first dataset are smaller than $512\times 512$, we do not include results for these two cubes.

The measurements are captured with $K=2$ shots, and the measurement rate is $m/n=KM(N+L+1)/(MNL)\approx0.069$ for ``natural scene 2002" and $0.065$ for ``natural scene 2004."
We test for measurement noise levels such that the SNRs are 15 dB and 20 dB. 
The typical runtimes for AMP with 400 iterations, GPSR with 400 iterations, and TwIST with 200 iterations are approximately $2,800$ seconds.
The average PSNR over all spectral bands for each reconstructed cube is shown in Tables~\ref{tb:scene2002} and~\ref{tb:scene2004}. We highlight the highest PSNR among AMP-3D-Wiener, GPSR, and TwIST using bold fonts.  It can be seen from Tables~\ref{tb:scene2002} and~\ref{tb:scene2004} that AMP-3D-Wiener usually outperforms GPSR by $2-5$ dB in terms of the PSNR, and outperforms TwIST by $0.2- 4$ dB, while TwIST outperforms GPSR by up to 3 dB for most of the scenes.
Additionally, the results of 6 selected image cubes are displayed in Figure~\ref{fig.scene} in the form of 2D RGB images.\footnote{We refer to the tutorial from http://personalpages.manchester.ac.uk/staff/\\david.foster/Tutorial$\_$HSI2RGB/Tutorial$\_$HSI2RGB.html and convert 3D image cubes to 2D RGB images.}
The four rows of images correspond to ground truth, results by AMP-3D-Wiener, results by GPSR, and results by TwIST, respectively. We can see from Figure~\ref{fig.scene} that 
the test datasets contain both smooth scenes and scenes with large gradients, and
AMP-3D-Wiener consistently reconstructs better than GPSR and TwIST, which suggests that AMP-3D-Wiener is adaptive to various types of scenes.

\begin{table}[t]
\vspace*{-0mm}
\centering
\begin{tabular}{|c|| c | c | c ||c|c|c|}
\hline
SNR&\multicolumn{3}{c||}{15 dB}&\multicolumn{3}{c|}{20 dB}\\
\hline
Algorithm& AMP & GPSR & TwIST & AMP & GPSR & TwIST \\
\hline 
Scene 1& {\bf32.69} & 28.10 & 31.05 & {\bf33.29} & 28.09 & 31.16\\

Scene 2& {\bf26.52} & 24.32 & 26.25 & {\bf26.65} & 24.40 & 26.41\\

Scene 3& {\bf32.05} & 29.33 & 31.21 & {\bf32.45} &  29.55 & 31.54\\
Scene 4& {\bf27.57} & 25.19 & 27.17 & {\bf27.76} & 25.47 &  27.70\\
Scene 5& {\bf29.68} & 27.09 & 29.07 & {\bf29.80} & 27.29 & 29.42\\
Scene 8& {\bf28.72} & 25.53 & 26.24 & {\bf29.33} & 25.77 & 26.46 \\
\hline
\end{tabular}     
\caption{\small\sl 
Average PSNR comparison of AMP-3D-Wiener, GPSR, and TwIST for the dataset ``natural scene 2002" downloaded from~\cite{Foster2006}. The spatial dimensions of the cubes are cropped to $M=N=512$, and each cube has $L=31$ spectral bands. 
The measurements are captured with $K=2$ shots, and the number of measurements is $m=557,056$.
Random Gaussian noise is added to the measurements such that the SNR is 15 or 20 dB. Because the spatial dimensions of the cubes ``scene 6" and ``scene7" in ``natural scenes 2002" are smaller than $512\times 512$, we do not include results for these two cubes.}
\label{tb:scene2002}
\end{table}

\begin{table}[t]
\vspace*{-0mm}
\centering
\begin{tabular}{|c|| c | c | c ||c|c|c|}
\hline
SNR&\multicolumn{3}{c||}{15 dB}&\multicolumn{3}{c|}{20 dB}\\
\hline
Algorithm& { AMP} & GPSR & TwIST & { AMP} & GPSR & TwIST \\
\hline 
Scene 1& {\bf30.48} & 28.43 &30.17 & {\bf30.37} & 28.53 & 30.31 \\

Scene 2& {\bf27.34} & 24.71 &27.03 & {\bf27.81} &  24.87 &  27.35 \\

Scene 3& {\bf33.13} & 29.38 & 31.69 & {\bf33.12} & 29.44 & 31.75\\
Scene 4& {\bf32.07} & 26.99 & 31.69 & {\bf32.14} & 27.25 & 32.08\\
Scene 5& {\bf27.44} & 24.25 & 26.48 & {\bf27.83} & 24.60 & 26.85\\
Scene 6& {\bf29.15} & 24.99 & 25.74 & {\bf30.00} & 25.53 & 26.15 \\
Scene 7& {\bf36.35} & 33.09 & 33.59 & {\bf37.11} & 33.55 & 34.05\\
Scene 8& {\bf32.12} & 28.14& 28.22 & {\bf32.93} & 28.82 & 28.69 \\
\hline
\end{tabular}     
\caption{\small\sl 
Average PSNR comparison of AMP-3D-Wiener, GPSR, and TwIST for the dataset ``natural scene 2004" downloaded from~\cite{Foster2006}. The spatial dimensions of the cubes are cropped to $M=N=512$, and each cube has $L=33$ spectral bands. 
The measurements are captured with $K=2$ shots, and the number of measurements is $m=559,104$.
Random Gaussian noise is added to the measurements such that the SNR is 15 or 20 dB.}
\label{tb:scene2004}
\end{table}

\begin{figure*}[t!]
\setcounter{cnt04}{9}
\vspace*{0mm}
\begin{center}
\includegraphics[width=160mm]{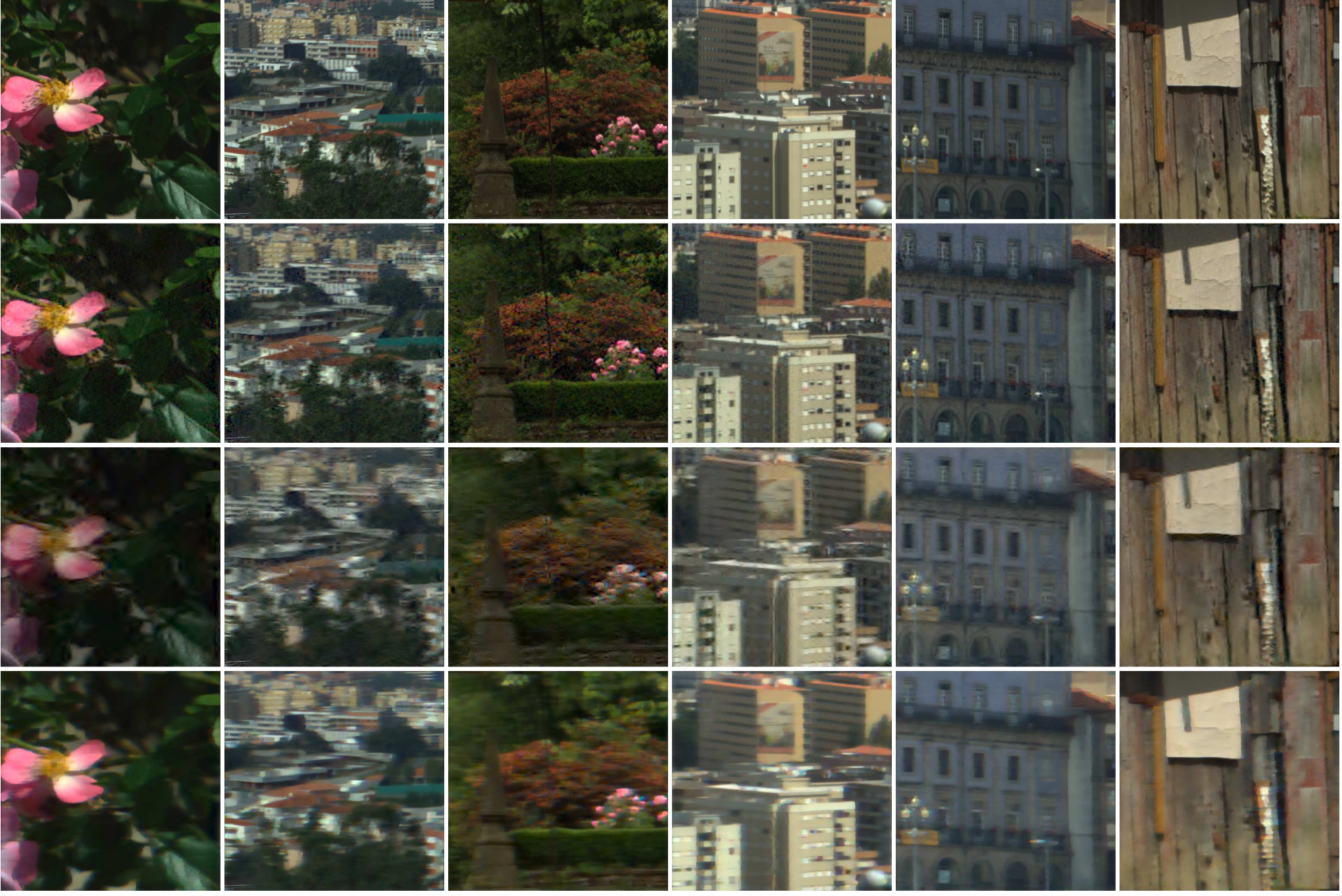}
\end{center}
\vspace*{0mm}
\caption{\small\sl 
Comparison of selected image cubes reconstructed by AMP-3D-Wiener, GPSR, and TwIST for the datasets ``natural scene 2002" and ``natural scene 2004." The 2D RGB images shown in this figure are converted from their corresponding 3D image cubes. Cube size is $N=M=512$, and $L=31$ for images in columns $1-2$ or $L=33$ for images in columns $3-6$.  Random Gaussian noise is added to the measurements such that the SNR is 20 dB. First row: ground truth; second row: the reconstruction result by AMP-3D-Wiener; third row: the reconstruction result by GPSR; last row: the reconstruction result by TwIST.}
\label{fig.scene}
\setcounter{figure}{\value{cnt04}}
\end{figure*}

\section{Conclusion}
\label{sec:disc}

In this paper, we considered the compressive hyperspectral imaging reconstruction problem for the coded aperture snapshot spectral imager (CASSI) system.
Considering that the CASSI system is a great improvement in terms of imaging quality and acquisition speed over conventional spectral imaging techniques, it is desirable to further improve CASSI by accelerating the 3D image cube reconstruction process. Our proposed AMP-3D-Wiener used an adaptive Wiener filter as a 3D image denoiser within the approximate message passing (AMP)~\cite{DMM2009} framework. AMP-3D-Wiener was faster than existing image cube reconstruction algorithms, and also achieved better reconstruction quality.

In AMP, the derivative of the image denoiser is required, and the adaptive Wiener filter can be expressed in closed form using a simple formula, and so its derivative is easy to compute.
Although the matrix that models the CASSI system is ill-conditioned and may cause AMP to diverge, we helped AMP converge using damping, and reconstructed 3D image cubes successfully.
Numerical results showed that AMP-3D-Wiener is robust and fast, and outperforms gradient projection for sparse reconstruction (GPSR) and two-step iterative shrinkage/thresholding (TwIST) even when the regularization parameters for GPSR and TwIST are optimally tuned.  Moreover, a significant advantage over GPSR and TwIST is that AMP-3D-Wiener need not tune any parameters, and thus an image cube can be reconstructed by running AMP-3D-Wiener only once,
which is critical in real-world scenarios. 
In contrast, GPSR and TwIST must be run multiple times in order to find the optimal regularization parameters. 

{\bf Future improvements: }In our current AMP-3D-Wiener algorithm for compressive hyperspectral imaging reconstruction, we estimated the noise variance of the noisy image cube within each AMP iteration using~\eqref{eq:sigma_t}. In order to denoise the noisy image cube in the sparsifying transform domain, we applied the estimated noise variance value to all wavelet subbands. The noise variance estimation and 3D image denoising method were effective, and helped produce promising reconstruction. However, both the noise variance estimation and the 3D image denoising method may be sub-optimal, because the noisy image cube within each AMP iteration does not contain i.i.d. Gaussian noise, and so the coefficients in the different wavelet subbands may contain different amounts of noise. 
On the other hand, in the proposed adaptive Wiener filter, the variances of the coefficients in the sparsifying transform domain were estimated empirically within each wavelet subband, whereas it is also possible to apply Wiener filtering via marginal likelihood or generalized cross validation~\cite{Rasmussen06}.
Therefore, it is possible that the denoising part of the proposed algorithm can be further improved. The study of such denoising methods is left for future work.

In our current AMP-3D-Wiener, the coded apertures must be complementary, because complementary coded apertures ensure that the norm of each column in the matrix~${\bf H}$ in~\eqref{eq:CASSI} is similar, otherwise, AMP-3D-Wiener may diverge. Although using complementary coded aperture has practical importance, it provides more flexibility in coded aperture design when such a complementary constraint can be removed, and the development for AMP-based algorithms without such constraints is left for future work.

Finally, besides reconstructing image cubes from compressive hyperspectral imaging systems, it would also be interesting to investigate problems such as target detection~\cite{Manolakis2003} and unmixing~\cite{Bioucas2012} using compressive measurements from hyperspectral imaging systems. We leave these problems for future work.

\section*{Acknowledgments}
We thank Sundeep Rangan and Phil Schniter for inspiring discussions on approximate message passing; Lawrance Carin, and Xin Yuan for kind help on numerical experiments; Junan Zhu for informative explanations about CASSI systems; Nikhil Krishnan for detailed suggestions on the manuscript; and the reviewers for their careful evaluation of the manuscript.

\ifCLASSOPTIONcaptionsoff
\newpage
\fi
\bibliographystyle{IEEEtran}
\bibliography{IEEEabrv,cites}

\end{document}